\documentclass{elsart}
\usepackage[mathscr]{euscript}

\usepackage{amssymb}

\newcommand*{\be}{\begin{equation}}
\newcommand*{\ee}{\end{equation}}
\newcommand*{\ba}{\begin{array}}
\newcommand*{\ea}{\end{array}}
\newcommand*{\bea}{\begin{eqnarray}}
\newcommand*{\eea}{\end{eqnarray}}
\newcommand*{\bean}{\begin{eqnarray*}}
\newcommand*{\eean}{\end{eqnarray*}}

\newcommand*{\lp}{\left(}
\newcommand*{\rp}{\right)}
\newcommand*{\ls}{\left[}
\newcommand*{\rs}{\right]}
\newcommand*{\lc}{\left\{}
\newcommand*{\rc}{\right\}}

\newcommand*{\la}{\langle}
\newcommand*{\La}{\left\la}
\newcommand*{\ra}{\rangle}
\newcommand*{\Ra}{\right\ra}

\renewcommand*{\d}{d}
\newcommand*{\g}{\mbox{\slshape g}}
\newcommand*{\gb}%
	{\mbox{\bf\slshape g%
	}}
\newcommand*{\p}{\partial}

\newcommand*{\Imm}{\,\Im{\mathfrak{m}}\,}
\newcommand*{\Ree}{\,\Re{\mathfrak{e}}\,}

\newcommand*{\veps}{\varepsilon}

\newcommand*{\mPi}{{\mathit\Pi}}

\newcommand*{\Hvf}{H_{\mathrm{vf}}}
\newcommand*{\Hgh}{H_{\mathrm{gh}}}

\newcommand*{\vf}{_{\mathrm{vf}}}
\newcommand*{\gh}{_{\mathrm{gh}}}
\renewcommand*{\pf}{_{\mathrm{f}}}

\newcommand*{\kB}{k_{\rm B}}

\newcommand*{\Tr}{\mathop{{\rm Tr}}\nolimits\,}

\newcommand*{\ddt}{\frac{\partial}{\partial t}}

\newcommand*{\ds}{\displaystyle}

\newcommand*{\APNY}{Ann. Phys. (NY)\ }
\newcommand*{\CMP}{Commun. Math. Phys.\ }
\newcommand*{\CMPL}{Cond. Matt. Phys.\ }

\newcommand*{\JCP}{J. Chem. Phys.\ }

\newcommand*{\JMP}{J. Math. Phys.\ }
\newcommand*{\JPA}{J. Phys. A\ }

\newcommand*{\JPSJ}{J. Phys. Soc. Japan\ }

\newcommand*{\MPLB}{Mod. Phys. Lett. B\ }

\newcommand*{\PA}{Physica A\ }

\newcommand*{\PRev}{Phys. Rev.\ }
\newcommand*{\PRA}{Phys. Rev. A\ }

\newcommand*{\PRD}{Phys. Rev. D\ }

\newcommand*{\PRL}{Phys. Rev. Lett.\ }
\newcommand*{\PLA}{Phys. Lett. A\ }
\newcommand*{\PLB}{Phys. Lett. B\ }

\newcommand*{\PTP}{Progr. Theor. Phys.\ }
\newcommand*{\PTPS}{Progr. Theor. Phys. Suppl.\ }

\newcommand*{\RMaP}{Rev. Math. Phys.\ }
\newcommand*{\RMoP}{Rev. Mod. Phys.\ }
\newcommand*{\ZP}{Z. Phys.\ }

\newlength{\ddash}
\providecommand*{\dbar}{\d\hspace*{-0.75\ddash}\rule[1.3ex]{0.4em}{0.05ex}}

\providecommand*{\e}{\textrm{e}}

\newcommand*{\ad}{a^{\dag}}
\newcommand*{\ta}{\tilde{a}}

\newcommand*{\tad}{\tilde{a}^{\dag}}
\newcommand*{\add}{a^{\dag\!\dag}}
\newcommand*{\tadd}{\tilde{a}^{\dag\!\dag}}

\newcommand*{\venus}{{\scriptscriptstyle+}\hspace{-1.5mm}^{\circ}}
\newcommand*{\tgdd}{\tilde{\gamma}^{\venus}}
\newcommand*{\tg}{\tilde{\gamma}}
\newcommand*{\gdd}{\gamma^{\venus}}

\begin{document}

\makeatletter
\@addtoreset{equation}{section}
\def\theequation{\arabic{section}.\arabic{equation}}
\makeatother

\begin{frontmatter}



\title{Quantum stochastic differential equations for boson and fermion systems
--- Method of Non-Equilibrium Thermo Field Dynamics}


\author{A.E. Kobryn\thanksref{K}}, \author{T. Hayashi} and 
\author{T. Arimitsu\corauthref{cor1}}\ead{arimitsu@cm.ph.tsukuba.ac.jp}
\address{Institute of Physics, University of Tsukuba, Ibaraki 305-8571, Japan}
\thanks[K]{Present address: Institute for Molecular Science, 
Myodaiji, Okazaki 444-8585, Japan}
\corauth[cor1]{Corresponding author}

\begin{abstract}
A unified canonical operator formalism for quantum stochastic differential equations,
including the quantum stochastic Liouville
equation and the quantum Langevin equation both of the It\^o
and the Stratonovich types,
is presented within the framework of Non-Equilibrium Thermo Field Dynamics (NETFD).
It is performed by introducing
an appropriate martingale operator in the Schr\"odinger and 
the Heisenberg representations with fermionic and bosonic Brownian motions.
In order to decide the double tilde conjugation rule and the thermal state conditions
for fermions, a generalization of the system consisting of a vector field
and Faddeev-Popov ghosts to dissipative open situations is carried out within
NETFD.
\end{abstract}

\end{frontmatter}

\section{Introduction}
\label{Introduction}

In this paper we study time-dependent behavior of non-equilibrium
quantum systems involving stochastic forces which can be boson or
fermion type and are called quantum Brownian motion. Present
consideration is an extension of previous analysis reported
comprehensively by one of the authors \cite{Arimitsu94} and is
given in terms of Non-Equilibrium Thermo Field Dynamics (NETFD)
\cite{Arimitsu85,Arimitsu87a,Arimitsu87b}. NETFD is a unified
formalism, which enables us to treat dissipative quantum systems
by the method similar to usual quantum mechanics or quantum
field theory, which accommodates the concept of the dual structure
in the interpretation of nature, i.e. in terms of the operator
algebra and the representation space. The representation space in
NETFD is composed of a direct product of two Hilbert spaces: 
one is for non-tilde fields, and the other for tilde fields. Within
the statistical operator (density operator) formalism there is entanglement between
operators and statistical operator due to their
non-commutativity. Introduction of two kinds of operators,
\textit{without} tilde and \textit{with} tilde, made it possible to
resolve the entanglement between relevant operators and the
statistical operator.

We are deriving a unified system of quantum stochastic differential equations
(QSDEs) under the influence of quantum Brownian motion, including the quantum stochastic
Liouville equation and the quantum Langevin equation. The quantum
Fokker-Planck equation is derived by taking the random average of
the corresponding stochastic Liouville equation. The relation
between the Langevin equation and the stochastic Liouville
equation, as well as between the Heisenberg equation for operators
of gross variables and the quantum Fokker-Planck equation obtained here,
is similar to the one between the Heisenberg equation and the
Schr\"o\-din\-ger equation in quantum mechanics and field theory.
Our extension of analysis \cite{Arimitsu94} consists of
essentially three items. Two of them include definition of fermionic
Brownian motion and treatment of fermions in NETFD, 
i.e. the tilde conjugation rule and the thermal state
conditions in the case of fermion systems. Third item is the
simultaneous consideration of hermitian and non-hermitian
interaction Hamiltonians.

To begin with, we first remind briefly
some standard steps that people usually take in order to obtain
the irreversible evolution of macroscopic systems starting from
the microscopic level.
At present, there are many viewpoints giving us tools how to
describe $N$-body systems out of equilibrium. At the same time,
one usually follows one of several basic approaches: (i) the
behavior of the systems is expressed in terms of not the total
($N$-particle) distribution function but $s$-particle ones (with
$s$ being usually 1 and/or 2), (ii) the dynamics of the systems is
characterized by the evolution of a ``coarse grained'' phase-space
distribution function or statistical operator, 
and (iii) the evolution of the systems is described by the
equations of motion for the dynamical gross variables.

The approach (i) is intimately related to the Bogoliubov
method of a reduced description of many-particle systems
\cite{Bogoliubov62}, which is widely used for construction of
kinetic equations based on the Liouville or the
Liouville-von-Neumann equation. Bogoliubov's hypothesis that the
time dependence of higher-particle distribution functions enter
through the one-particle distribution provides a fundamental
importance in various schemes of truncation of the BBGKY
hierarchy.

In the approach (ii), the most frequently used tools are projection
operators introduced by Nakajima \cite{Nakajima58} and Zwanzig
\cite{Zwanzig60a,Zwanzig60b}. The basic idea underlying the
application of their techniques to complex systems is to regard
the operation of tracing over the environment as a formal
projection in the space of the total system. It became especially
popular in quantum optics where the so-called quantum master equation 
for reduced statistical operator of a relevant system 
now bears their names and is
called the Nakajima-Zwanzig equation \cite{Breuer02}.

The general framework, called sub-dynamics, at the Brussels school 
is also related to the
approach (ii) but the underlying concept is different from the one by
Nakajima and Zwanzig. The main point here is the notion of the increase of the
number of correlations within a system in time. It has been expounded in detail by
Prigogine and coauthors, see e.g. \cite{Prigogine69,Prigogine99}.

Regarding to the approach (iii), we should mention projection
operator by Mori \cite{Mori65} and the one by Kawasaki and Gunton
\cite{Kawasaki73}. The former is used to derive linear equations
of motion for gross variables out of non-linear equations.
Originally, one of the intentions to introduce such an operator
was to obtain expressions for physical (measurable) kinetic
coefficients. The latter is an improved version of the
time-dependent projection operator by Robertson
\cite{Robertson66}.

Zubarev introduced the concept of \textit{non-equilibrium
ensemble} as a generalization of Green's works on the statistical
mechanics of linear dissipation processes \cite{Green52,Green54}
and Kubo's theory of linear response of systems to
mechanical \cite{Kubo57a} and thermodynamical \cite{Kubo57b}
external perturbations. This generalization is known as the method
of non-equilibrium statistical operator \cite{Zubarev74}. It is shown that
this has a close relationship to the projection operator methods \cite{Zubarev81}.

Above mentioned methods do not exhaust the entire list, but they
may be the most generic ones. However, in this paper we do not follow
them. In the case of presence of additional degree(s) of freedom,
e.g. stochastic force(s), description may be given also in some
optional way (in a sense that consideration does not start from
the very microscopic level). The theory of Brownian motion is an
example. The fundamental equation here is the Langevin
equation and it is the stochastic differential equation for
dynamical variables \cite{Chandrasekhar43,Wang45}. Random forces
in Langevin equation are usually described by Gaussian white
stochastic processes. Stochastic integral with respect to such
processes is defined as a kind of a Riemann-Stieltjes one
\cite{Gardiner85book} where multiplication between the stochastic
increment and integrand is commonly considered in the form of
It\^o~\cite{Ito44} or Stratonovich~\cite{Stratonovich66} (for
It\^o and Stratonovich multiplications see Appendix \ref{IS}).

The Langevin equation can be used to calculate various time
correlation functions. Now it is radically extended to solve
numerous problems in different areas
\cite{Shuler69,Coffey85,Sobczyk91,vanKampen92,Coffey96}. In
particular, the theory of Brownian motion itself has been extended
to situations where the ``Brownian particle'' is not a real
particle anymore, but instead some collective properties of a
macroscopic system. Corresponding equation in the phase space or
the Liouville space of statistical operators
can be considered as a sort of stochastic differential equation
too. In order to investigate classical stochastic systems, the
stochastic Liouville equation was introduced first by Anderson
\cite{Anderson54} and Kubo \cite{Kubo54,Kubo62,Kubo63}.

There were several attempts to extend the classical theory (both
the Langevin and the stochastic Liouville equations) for quantum cases.
Study of the Langevin equation for quantum systems has its origin
in papers by Senitzky \cite{Senitzky60,Senitzky61,Senitzky63},
Schwinger \cite{Schwinger61}, Haken
\cite{Haken64,Haken65,Haken70,Haken75} and Lax \cite{Lax66}, where
they investigated a quantum mechanical damped harmonic oscillator
in connection with laser systems. In particular, it was shown that
the quantum noise, i.e. the spontaneous emission, can be treated
in a way similar to the thermal fluctuations, and that the noise
source has non-zero second moments proportional to a quantity
which can be associated with a quantum analog of a diffusion
coefficient. As it was noticed by Kubo \cite{Kubo69} in his
discussion with van~Kampen, the random force must be an operator
defined in its own Hilbert space, which does not happen in
classical case since there is no consideration of space for the
random force.

Mathematical study of the quantum stochastic processes was
initiated by Davies \cite{Davies69,Davies76}, Hudson
\cite{Cockroft77,Hudson81,Hudson84a,Hudson84b,Hudson84c,Hudson85},
Accardi \cite{Accardi82,Accardi90}, Parthasarathy
\cite{Parthasarathy85,Parthasarathy89,Parthasarathy92} and their
co-authors. Quantum mechanical analogs of Wiener processes
\cite{Cockroft77} and quantum It\^o formula for boson systems
\cite{Hudson81,Hudson84a,Hudson84b,Hudson84c} were defined first
by Hudson \textit{et al}.. The classical Brownian motion is
replaced here by the pair of one-parameter unitary group
authomorphisms, namely by the annihilation and creation boson
random force operators with time indices in the boson Fock space,
named quantum Brownian motion. Fermion stochastic calculus were
defined by Applebaum, Hudson and Parthasarathy
\cite{Applebaum84a,Applebaum84b,Applebaum86,Applebaum95,Hudson86,Parthasarathy86}.
In these papers, they developed the fermion analog of the
corresponding boson theory \cite{Hudson84a,Hudson84b} in which the
annihilation and creation processes are fermion field operators in
the fermion Fock space. Within the frame of this formalism, the
It\^o-Clifford integral
\cite{Barnett82,Barnett83a,Barnett83b,Barnett83c}-- fermion analog
of the classical Brownian motion -- is contained as a special
case. It should be noted, however, that in both boson and fermion
theories of quantum stochastic calculus mathematicians were
debating unitary processes only. For readers' convenience, clue of
mathematicians' theory of quantum Brownian motion and the
extension with allowance for thermal degree of freedom are put
into Appendix \ref{bfbm}.

Contrary to expectations, attempts to extend the classical
stochastic Liouville equation for quantum case were not very
successful so far. In present work we construct our consideration
using the formalism of NETFD. It is an alternative way to the
above mentioned general methods of non-equilibrium statistical
mechanics in the sense that it provides us with a general
structure of the canonical operator formalism for dissipative
non-equilibrium quantum systems without starting from the
microscopic description, and turns out to be especially successful
in the inclusion of quantum stochastic forces. In particular, a
unified canonical operator formalism of QSDEs 
for boson systems was constructed first within NETFD
\cite{Arimitsu91a,Arimitsu91b,Saito92,Saito93a,Saito93b,Saito97,Arimitsu94}
on the basis of the quantum stochastic Liouville equation.

The paper is organized as follows. First, in section
\ref{Basics-of-NETFD}, we remind a brief essence of the formalism
of  NETFD by giving its technical basics and some fundamentals. In
section \ref{Semi-free-Ht} we derive the semi-free time evolution
generator for systems in non-stationary case. The semi-free
generator is bi-linear  and globally gauge invariant. The
annihilation and creation operators are introduced by means of a
time-dependent Bogoliubov transformation. We close the section by
calculating the two-point function. The generating functional
method, which gives us the relation between the method of
NETFD and the one of the Schwinger closed-time path, is introduced
in section \ref{GF-method}. Interaction with external fields is
considered in section \ref{Interaction}. Here we study two cases:
hermitian and non-hermitian interaction hat-Hamiltonians.
To make possible their
si\-mul\-ta\-ne\-o\-us consideration we 
in\-tro\-du\-ce an auxiliary parameter $\lambda$ which plays the
r\^ole of a switch between the cases. In section \ref{QSDE}, the
general expression of the stochastic semi-free time evolution
generator is derived for a non-stationary Gaussian white quantum
stochastic process by means of the interaction hat-Hamiltonian
with arbitrary $\lambda$. Correlations of the random force
operators are also derived generally. With the generator, quantum
stochastic Liouville equations and quantum stochastic Langevin
equations of both It\^o and Stratonovich types of the system are
investigated in a unified manner. We conclude the section by
deriving the equation of motion for the expectation value of an
arbitrary operator of the relevant system. In section
\ref{Stationary}, we consider a semi-free system with a stationary
process and check explicitly the irreversibility of such a process
in terms of its Boltzmann entropy. In section \ref{MC-WF} we
investigate relation to the Monte Carlo wave-function method.
Summary and open questions are put into section
\ref{Conclusions}. Auxiliary material is put into Appendices.

\section{Basics of NETFD}
\label{Basics-of-NETFD}

Information about the general method of NETFD can be found in many
papers and we refer first of all to the original source
\cite{Arimitsu85,Arimitsu87a,Arimitsu87b} and the review article
\cite{Arimitsu94}. To make our paper self-contained, we include
some standard steps which are necessary at least to fix the
notations. The formalism of NETFD is constructed upon the
following fundamental requirements.

An arbitrary operator $A$ in NETFD is accompanied by its
\textit{tilde conjugated partner} $\tilde{A}$, called tilde
operator, according to the rule
%
\bea
(A_1A_2)^\sim&=&\tilde{A}_1\tilde{A}_2,\label{tc1}\\
(c_1A_1+c_2A_2)^\sim&=&c_1^*\tilde{A}_1+c_2^*\tilde{A}_2,\label{tc2}\\
(\tilde{A})^\sim&=&A,\label{tc3}
\eea
%
where $c_1$ and $c_2$ are $c$-numbers. It should be noted that in
the present paper the double tilde conjugation rule (\ref{tc3}) is
of the same form for both bosonic and fermionic operators and
leaves them unchanged.

To indicate commutation or anti-commutation of two operators, say
$A_1$ and $A_2$, we will use the notation $[ A_1,A_2 \}$ and call it
(anti-)\-com\-mu\-ta\-tor, which should be understood as
\bea
[ A_1,A_2 \} =[A_1,A_2]_+=\{A_1,A_2\}=A_1A_2+A_2A_1
\label{sigma-commutator1}
\eea
when both operators are fermionic, or
\bea
[ A_1,A_2 \}=[A_1,A_2]_-=[A_1,A_2]=A_1A_2-A_2A_1
\label{sigma-commutator2}
\eea
otherwise.%
\footnote{When one operator is bosonic and another one is
fermionic, the rule of commutation depends on the system. In this
paper for such combinations we assume (\ref{sigma-commutator2}).}

Tilde and non-tilde operators, say $A_1$ and $\tilde{A}_2$, are
supposed to be mutually (anti-)\-com\-mu\-ta\-ti\-ve at equal
time, i.e.
\bea
[ A_1,\tilde{A}_2 \} =0.
\label{ctnt}
\eea

Tilde and non-tilde operators are related with each other through
the \textit{thermal state condition} (TSC)
\bea
\la\theta|\tilde{A}^\dag&=&\tau^*\la\theta|A,\label{tsc}
\eea
where $\la\theta|$ represents the \textit{thermal bra-vacuum};
$\tau$ is the \textit{complex} parameter which takes two values:
\bea
\tau&=&\lc
\ba{ccl}
1&\;&\mbox{for bosonic operators,}\\
i&&\mbox{for fermionic operators.}\\
\ea
\right.
\label{tau}
\eea
Derivation of the double tilde conjugation rule and TSC for
fermionic operators used in this paper is given in
Ap\-pen\-dix~\ref{phase}.

Within the framework of NETFD, the dynamical evolution of a system
is described by the Schr\"o\-din\-ger equation (here we use the
system with $\hbar=1$)
\bea
\ddt|\mathit{0}(t)\ra&=&-i\hat{H}|\mathit{0}(t)\ra,
\label{Schrodinger-eq}
\eea
where $|\mathit{0}(t)\ra$ represents the \textit{thermal
ket-vacuum}. 
It can be also called the quantum master equation or 
the quantum Fokker-Planck equation in this paper.
The thermal vacuums are tilde invariant, i.e.,
$\la\theta|^\sim=\la\theta|$ and
$|\mathit{0}(t)\ra^\sim=|\mathit{0}(t)\ra$, and are normalized as
$\la\theta|\mathit{0}(t)\ra=1$. The hat-Hamiltonian $\hat{H}$, an
infinitesimal time-evolution generator, satisfies the
\textit{tildian} condition:
\bea
(i\hat{H})^\sim&=&i\hat{H}.\label{tildian}
\eea
The tildian hat-Hamiltonian is not necessarily hermitian operator.
It has zero eigenvalues for the thermal bra-vacuum
\bea
\la\theta|\hat{H}&=&0,\label{zeroev}
\eea
which is nothing but manifestations of conservation of probability.

Introducing the time-evolution operator $\hat{V}(t)$ by
\bea
\frac{\d}{\d t}\hat{V}(t)&=&-i\hat{H}\hat{V}(t)
\eea
with the initial condition $\hat{V}(0)=1$, we can define the
Heisenberg operator
\bea
A(t)&=&\hat{V}^{-1}(t)A\hat{V}(t)
\eea
satisfying the Heisenberg equation for dissipative systems
\bea
\frac{\d}{\d t}A(t)&=&i[\hat{H}(t),A(t)],\label{HEDS}
\eea
where $\hat{H}(t)$ is the hat-Hamiltonian in the Heisenberg
representation. The existence of the Hei\-sen\-berg equation of
motion for coarse grained operators enables us to construct a
canonical formalism of the dissipative quantum field theory. Note
that with the help of TSC we have an equation of motion for a
vector $\la\theta|A(t)$
\bea
\frac{\d}{\d t}\la\theta|A(t)&=&i\la\theta|[\hat{H}(t),A(t)]
\eea
in terms of only non-tilde operators. The expectation value of an
observable operator $A$ at time $t$ is given by
\bea
\la A(t)\ra&=&\la\theta|A|\mathit{0}(t)\ra
=\la\theta|A(t)|\mathit{0}\ra,
\eea
where $|\mathit{0}\ra=|\mathit{0}(t=0)\ra$. We define that
observable operators consist only of non-tilde operators.%
\footnote{We can include tilde operators in addition to non-tilde
ones in the definition of observable. However, inclusion of tilde
operators may give us a set of different but equivalent definitions for
\textit{one} observable operator.}

\section{Semi-free hat-Hamiltonian}
\label{Semi-free-Ht}

Let us consider a system specified by the total hat-Hamiltonian
\bea
\hat{H}^{\mathrm{tot}}_t&=&\hat{H}_t+\hat{H}_1+
\hat{H}_{\mathrm{I},t},\label{H-total}
\eea
where $\hat{H}_t$ is a semi-free hat-Hamiltonian, whereas
$\hat{H}_1$ and $\hat{H}_{\mathrm{I},t}$ are, respectively,
the interaction hat-Hamiltonian within the relevant system and the one
representing the coupling with external fields.
The system
itself is supposed to be consistent with all the requirements of
NETFD given in the previous section. 
Some general remark about treatment of interaction within the
relevant system is given in section \ref{GF-method}. Explicit treatment of
interaction with external fields is given in section
\ref{Interaction}. Here we concentrate on derivation and study of
properties of the semi-free hat-Hamiltonian, i.e. renormalized
unperturbed hat-Hamiltonian.

\subsection{Derivation of the semi-free hat-Hamiltonian}

The semi-free hat-Hamiltonian is bilinear in operators $a$, $\ad$,
$\ta$ and $\tad$, and is invariant under the phase transformation
$a\to a\e^{i\phi}$:
\bea
\hat{H}_t=h_1(t)\ad a+h_2(t)\tad\ta+h_3(t)a\ta+h_4(t)\ad\tad+h_0(t),
\label{Ht-init}
\eea
where $h_j(t)$ are time-dependent complex $c$-number functions.
Operators $a$, $\ad$, $\ta$ and $\tad$ satisfy the canonical
(anti-)commutation relations
\bea
[a_{\mathbf k},\ad_{\mathbf k'}]_{-\sigma}&=&\delta_{{\mathbf k},\mathbf{k}'},
\quad
[\ta_{\mathbf k},\tad_{\mathbf k'}]_{-\sigma}=\delta_{{\mathbf k},\mathbf{k}'},
\eea
where we use $\sigma=1$ for bosonic systems and $\sigma=-1$ for
fermionic ones. According to (\ref{ctnt}), tilde and non-tilde
operators are mutually (anti-)com\-mu\-ta\-ti\-ve. In the
following account, a subscript $\mathbf{k}$ for specifying a
momentum and/or other degrees of freedom will be dropped unless it
is necessary. Number of unknown functions $h_j(t)$ can be reduced
by the use of (\ref{zeroev}) and tildian (\ref{tildian}) for the
semi-free hat-Hamiltonian. It results in
\bea
\hat{H}_t&=&\hat{H}_{\mathrm{S},t}+i\hat{\mPi}_t,
\label{Hs_Pi}
\eea
where
\bea
\hat{H}_{\mathrm{S},t}&=&\omega(t)\lp\ad a-\tad\ta\rp,\label{HSt}\\
\hat{\mPi}_t&=&c_1(t)\lp\ad a+\tad\ta\rp-\tau\ls2c_1(t)+c_2(t)\rs\ad\tad\nonumber\\
&&{}+\sigma\tau c_2(t)a\ta+\sigma\ls2c_1(t)+c_2(t)\rs,\label{mPit1}\qquad
\eea
with
%
\bea
\omega(t)&=&\Ree h_1(t),\\
c_1(t)&=&\Imm h_1(t),\\[2ex]
c_2(t)&=&\lc
\ba{ll}
\Imm h_3(t)&\mbox{for bosonic systems,}\\
\Ree h_3(t)&\mbox{for fermionic systems.}\\
\ea\right.
\eea
%

Let us introduce operators $a(t)$ and $\add(t)$ in the interaction
representation defined by
\bea
a(t)=\hat{V}^{-1}(t)a\hat{V}(t),\quad
\add(t)=\hat{V}^{-1}(t)\ad\hat{V}(t),
\eea
where
\bea
\label{hatv}
\frac{\d}{\d t}\hat{V}(t)&=&-i\hat{H}_t\hat{V}(t),
\eea
with the initial condition $\hat{V}(0)=1$. They satisfy the
equal-time (anti-)commutation relations
\bea
[a(t),\add(t)]_{-\sigma}&=&1.
\eea
The Heisenberg equation (\ref{HEDS}) for $a(t)$ and $\add(t)$ with
\bea
\hat{H}(t)=\hat{V}^{-1}(t)\hat{H}_t\hat{V}(t)
\eea
are explicitly given by
%
\bea
\frac{\d a(t)}{\d t}&=&\ls c_1(t)-i\omega(t)\rs a(t)-
\tau\ls2c_1(t)+c_2(t)\rs\tadd(t),
\label{Heisenberg-eq0}\\
\frac{\d\add(t)}{\d t}&=&\ls i\omega(t)-c_1(t)\rs\add(t)-\tau c_2(t)\ta(t).
\label{Heisenberg-eq}
\eea
%
In these formulae we used a symbol $\dag\!\dag$ instead of usual
dagger because the semi-free hat-Hamiltonian $\hat{H}_t$ is not
necessarily hermitian.

Since the semi-free hat-Hamiltonian $\hat{H}_t$ satisfies
(\ref{zeroev}), we have TSC for the bra-vacuum at time $t$
\bea
\label{tsc-oir1}
\la\theta|\tadd(t)&=&\tau^*\la\theta|a(t).
\eea

By making use of the Heisenberg equations (\ref{Heisenberg-eq0}) 
and (\ref{Heisenberg-eq}), and of TSC (\ref{tsc-oir1}),
one obtains the equation of motion for a vector
$\la\theta|\add(t)a(t)$ in the form
\bea
\frac{\d}{\d t}\la\theta|\add(t)a(t)=
-2\kappa(t)\la\theta|\add(t)a(t)+i\Sigma^<(t)\la\theta|,
\label{eq-m}
\eea
where $\kappa(t)$ and $i\Sigma^<(t)$ are defined by
%
\bea
\kappa(t)&=&c_1(t)+c_2(t),
\label{kappa-Sigma0}\\
i\Sigma^<(t)&=&-\sigma[2c_1(t)+c_2(t)].
\label{kappa-Sigma}
\eea
%
Substituting (\ref{kappa-Sigma0}) and (\ref{kappa-Sigma}) into 
(\ref{Heisenberg-eq0}) and (\ref{Heisenberg-eq}) one
gets equations of motion for operators $a(t)$ and
$a^{\dag\!\dag}(t)$ in the form
%
\bea
\frac{\d a(t)}{\d t}&=&-\ls i\omega(t)+\kappa(t)\rs a(t)
-\sigma i\Sigma^<(t)\ls a(t)-\tau\tadd(t)\rs\!,\\
\frac{\d\add(t)}{\d t}&=&\ls i\omega(t)+\kappa(t)\rs\add(t)
+\sigma i\Sigma^<(t)\ls\add(t)-\tau\ta(t)\rs-2\tau\kappa(t)\ta(t).
\label{hem1}
\eea
%

Applying the thermal ket-vacuum $|\mathit{0}\ra$ at the initial
time to (\ref{eq-m}), we obtain the equation of motion for the
one-particle distribution function
\bea
n(t)&=&\la\theta|\add(t)a(t)|\mathit{0}\ra
\eea
in the form
\bea
\label{ndot}
\frac{\d}{\d t}n(t)&=&-2\kappa(t)n(t)+i\Sigma^<(t).
\eea
Equation (\ref{ndot}) can be identified as the generalized
Bolt\-zmann equation of the system. The function $i\Sigma^<(t)$ is
given when the interaction hat-Hamiltonian $\hat{H}_1$ is defined.

The initial ket-vacuum $|\mathit{0}\ra$ is specified by TSC
\bea
\tilde{a}|\mathit{0}\ra&=&\tau f\,a^\dag|\mathit{0}\ra\label{tsc2}
\eea
with $f\in\mathbf{R}$. The initial value for the one-particle
distribution function $n=n(t=0)$ is determined by $f$. Since
\bea
n&=&n^*=\la\theta|a^\dag a|\mathit{0}\ra^\sim
=\la\theta|\tilde{a}^\dag\tilde{a}|\mathit{0}\ra
=\tau^*\la\theta|a\tilde{a}|\mathit{0}\ra
=|\tau|^2f\la\theta|aa^\dag|\mathit{0}\ra\nonumber\\
&=&f[1+\sigma n],\label{n(f)1}
\eea
we have
\bea
n&=&f[1-\sigma f]^{-1}.\label{n(f)2}
\eea
In the first equality of (\ref{n(f)1}) we used the fact that $n$
is a real number; in the third equality we used the tilde
invariance of the thermal vacuums $\la\mathit{\theta}|$ and
$|\mathit{0}\ra$; finally, in the fourth and fifth equalities we
used TSCs (\ref{tsc}) and (\ref{tsc2}), respectively.

Solving the Heisenberg equations for $a(t)$, $\add(t)$ and their
tilde conjugates, and using TSC at initial time (\ref{tsc2}), we
find TSC for the ket-vacuum at time $t$
\bea
\label{tsc-oir2}
\tilde{a}(t)|\mathit{0}\ra&=&
\frac{\tau n(t)}{1+\sigma n(t)}\add(t)|\mathit{0}\ra,
\eea
where $n(t)$ satisfies the Boltzmann equation (\ref{ndot}).

Substituting (\ref{kappa-Sigma0}), (\ref{kappa-Sigma}) and (\ref{ndot}) into
(\ref{mPit1}) one gets the most general form of $\hat{\mPi}_t$
in the interaction representation:
\bea
\hat{\mPi}_t&=&-\lc\kappa(t)\ls1+2\sigma n(t)\rs+\sigma\dot{n}(t)\rc
\lp\ad a+\tad\ta\rp\nonumber\\
&&+\sigma\tau\lc2\kappa(t)\ls1+\sigma n(t)\rs+\sigma\dot{n}(t)\rc
a\tilde{a}\nonumber\\
&&+\sigma\tau\lc2\kappa(t)n(t)+\dot{n}(t)\rc\ad\tad-\lc2\kappa(t)n(t)+\dot{n}(t)\rc.
\label{Pit2}
\eea
Here, we used the abbreviation $\dot{n}(t)=\d n(t)/\d t$.

By introducing thermal doublet notations
\bea
\bar{a}^\nu=\lp\ad,-\tau\ta\rp,\quad
a^\mu={\mathrm{collon}}\lp a,\tau\tad\rp,
\label{tdn-a}
\eea
canonical (anti-)commutation relations are written as
\bea
[a^\mu,\bar{a}^\nu]_{-\sigma}&=&\delta^{\mu\nu}.
\eea
The resulting semi-free hat-Hamiltonian (\ref{Hs_Pi}) can be
presented in a compact form as
\bea
\hat{H}_t&=&\omega(t)\bar{a}^\mu a^\mu
+i\bar{a}^\mu A(t)^{\mu\nu}a^\nu
+\sigma[\omega(t)+i\kappa(t)],\qquad
\label{sfhh}
\eea
where matrix $A(t)^{\mu\nu}$ has the following structure:
\bea
A(t)^{\mu\nu}&=&\sigma\lp
\ba{lr}
\ds-\kappa(t)\ls2n(t)+\sigma\rs-\dot{n}(t),
&\ds2\kappa(t)n(t)+\dot{n}(t)\\
\ds-2\kappa(t)\ls n(t)+\sigma\rs-\dot{n}(t),
&\ds\kappa(t)\ls2n(t)+\sigma\rs+\dot{n}(t)
\ea
\rp\!.
\eea

\subsection{Annihilation and creation operators}

Let us introduce annihilation and creation operators by
%
\bea
\gamma(t)&=&\ls1+\sigma n(t)\rs a(t)-\sigma\tau n(t)\tadd(t),
\label{aco0}\\
\tgdd(t)&=&\tadd(t)-\sigma\tau a(t).
\label{aco1}
\eea
%
From TSCs (\ref{tsc-oir1}) and (\ref{tsc-oir2}) at time $t$, we see that
they annihilate the vacuums:
\bea
\label{gamma-annihilate}
\la\theta|\gdd(t)=0,\quad &&
\gamma(t)|\mathit{0}\ra=0,\\
\label{gamma-annihilate tilde}
\la\theta|\tgdd(t)=0,\quad &&
\tilde{\gamma}(t)|\mathit{0}\ra=0.
\eea
With the thermal doublet notations
\bea
\label{tdn-at}
\bar{a}(t)^\mu&=&\lp\add(t),-\tau\ta(t)\rp,\;\;
a(t)^\nu={\mathrm{collon}}\lp a(t),\tau\tadd(t)\rp,\\
\label{tdn-gt}
\bar{\gamma}(t)^\nu&=&\lp\gdd(t),-\tau\tg(t)\rp,\;\;
\gamma(t)^\mu={\mathrm{collon}}\lp\gamma(t),\tau\tgdd(t)\rp,
\eea
(\ref{aco0}), (\ref{aco1}) and their tilde conjugates can be written as
\bea
\bar{\gamma}(t)^\nu=\bar{a}(t)^\mu[B^{-1}(t)]^{\mu\nu},\quad
\gamma(t)^\mu=B(t)^{\mu\nu}a(t)^\nu,
\eea
where $B(t)^{\mu\nu}$ is a matrix of the time-dependent Bogoliubov
transformation:
\bea
B(t)^{\mu\nu}&=&\lp
\ba{cc}
1+\sigma n(t)&-\sigma n(t)\\
-1&1\\
\ea\rp.
\label{tdbt}
\eea
This transformation is the canonical one since it leaves the
canonical $(\mbox{anti-})$\-commutation relations unchanged:
\bea
[\gamma(t)^\mu,\bar{\gamma}(t)^\nu]_{-\sigma}&=&\delta^{\mu\nu}.
\eea

The equation of motion for the thermal doublet $\gamma(t)^\mu$ is
derived as
\bea
\label{eq-m-gammat}
\frac{\d}{\d t}\gamma(t)^\mu&=&\ls-i\omega(t)\delta^{\mu\nu}
-\kappa(t)\tau_3^{\mu\nu}\rs\gamma(t)^\nu,
\eea
where
\bea
\tau_3^{\mu\nu}=\lp
\ba{cr}
1&0\\
0&-1\\
\ea\rp.
\eea
The solution of (\ref{eq-m-gammat}) then is obtained in the form
\bea
\gamma(t)^\mu=\exp\lc\int_{0}^t\d t'\;\ls-i\omega(t')\delta^{\mu\nu}
-\kappa(t')\tau_3^{\mu\nu}\rs\rc\gamma(0)^\nu.
\eea

\subsection{Schr\"odinger representation}

Annihilation and creation operators in the Schr\"odinger representation
are introduced by the relations
\bea
\label{gamma-sr}
\bar{\gamma}(t)^\nu=\hat{V}^{-1}(t)\bar{\gamma}^\nu_t\hat{V}(t),\quad
\gamma(t)^\mu=\hat{V}^{-1}(t)\gamma^\mu_t\hat{V}(t),
\eea
with $\hat{V}(t)$ being specified by (\ref{hatv}) and the thermal
doublet notations
\bea
\label{tdn-g}
\bar{\gamma}^\nu_t=\lp\gdd,-\tau\tg_t\rp,\quad
\gamma^\mu_t={\mathrm{collon}}\lp\gamma_t,\tau\tgdd\rp.
\eea
Using (\ref{tdn-a}), one can write
\bea
\label{gamma-bt}
\bar{\gamma}^\nu_t=\bar{a}^\mu[B^{-1}(t)]^{\mu\nu},\quad
\gamma^\mu_t=B(t)^{\mu\nu}a^\nu,
\eea
where matrix $B(t)^{\mu\nu}$ is given by (\ref{tdbt}). 

We see that the annihilation and creation operators in the Schr\"odinger representation
annihilate the vacuums at time $t$:
\bea
\label{gamma-annihilate Sh}
\la\theta|\gdd =0,\quad &&
\gamma_t|\mathit{0}(t)\ra=0,\\
\label{gamma-annihilate tilde Sh}
\la\theta|\tgdd = 0,\quad &&
\tilde{\gamma}_t |\mathit{0}(t)\ra=0.
\eea
Note that creation operator $\gamma^{\venus}$ and its tilde
conjugated partner $\tilde{\gamma}^{\venus}$ do not depend on
time. It is consistent with the fact that the vacuum $\la\mathit{\theta}|$ does not
depend on time due to the property
$\la\mathit{\theta}|\hat{H}_t=0$.

The terms (\ref{HSt}) and (\ref{Pit2}) of the se\-mi-free
hat-Ha\-mil\-to\-ni\-an $\hat{H}_t$, (\ref{Hs_Pi}), now read
%
\bea
\hat{H}_{\mathrm{S},t}&=&\omega(t)\lp\gdd\gamma_t-\tgdd\tg_t\rp,
\label{3.43}\\
\hat{\mPi}_t&=&-\kappa(t)\lp\gdd\gamma_t+\tgdd\tg_t\rp
+\sigma\tau\dot{n}(t)\gdd\tgdd,\qquad\label{Pit3}
\eea
%
in the normal ordering with respect to the annihilation and creation operators
in the Schr\"odinger representation. 
When the system is semi-free, putting (\ref{3.43}) and (\ref{Pit3}) into
(\ref{Hs_Pi}) and substituting $\hat{H}_t$ for $\hat{H}$ into the
Schr\"odinger equation (\ref{Schrodinger-eq}), one has
\bea
\ddt|\mathit{0}(t)\ra&=&\sigma\tau\dot{n}(t)\gdd\tgdd|\mathit{0}(t)\ra.
\eea
It is solved to give
\bea
\label{ket1}
|\mathit{0}(t)\ra
&=&\exp\lc-\la\theta|\tilde{\gamma}_t\gamma_t
|\mathit{0}\ra\gamma^{\venus}\tilde{\gamma}^{\venus}\rc|\mathit{0}\ra.
\eea
Here we introduced a kind of the order parameter
\bea
\la\theta|\tilde{\gamma}_t\gamma_t|\mathit{0}\ra&=&\sigma\tau[n(0)-n(t)]
\label{order-parameter}
\eea
which gives a measure of difference of the system from the initial
state. From (\ref{ket1}) we see that the evolution of the ket-vacuum
is realized by a condensation of tilde and non-tilde particle pairs into
initial ket-vacuum. The ket-vacuum itself is the functional of the
one-particle distribution function $n(t)$. The dependence of the
thermal ket-vacuum on $n(t)$ is given by
\bea
\frac{\delta}{\delta n(t)}|\mathit{0}(t)\ra&=&
\sigma\tau\gdd\tgdd|\mathit{0}(t)\ra.
\eea
Then the Schr\"odinger equation can be written in an alternative way:
\bea
\lc\ddt-\dot{n}(t)\frac{\delta}{\delta n(t)}\rc|\mathit{0}(t)\ra&=&0.
\eea
This shows that the vacuum $|\mathit{0}(t)\ra$ is migrating in 
the super-representation space spanned by the one-particle distribution function
$\{ n_{\mathbf k}(t) \}$ with the {\it velocity} $\{ \dot{n}_{\mathbf k}(t) \}$
as a conserved quantity \cite{Arimitsu-essay96,Arimitsu-Waseda03}.

\subsection{Two-point function of the semi-free field}

A time-ordered two-point function $G(t,t')^{\mu\nu}$ (propagator),
defined by
\bea
G(t,t')^{\mu\nu}&=&-i\la\theta|T[a(t)^\mu\bar{a}(t')^\nu]|\mathit{0}\ra,
\eea
is given by
\bea
G(t,t')^{\mu\nu}&=& [B^{-1}(t)]^{\mu\lambda}{\mathcal{G}}(t,t')^{\lambda\rho}
B(t')^{\rho\nu},
\eea
with
\bea
{\mathcal{G}}(t,t')^{\lambda\rho}
=-i\la\theta|T[\gamma(t)^\lambda\bar{\gamma}(t')^\rho]|\mathit{0}\ra
=\lp
\ba{ll}
G^{\mathrm{R}}(t,t')&0\\
0&G^{\mathrm{A}}(t,t')\\
\ea\rp\label{Gmunu}
\eea
where non-zero matrix elements are
%
\bea
G^{\mathrm{R}}(t,t')&=&-i\la\theta|T[\gamma(t)\gdd(t')]|\mathit{0}\ra\nonumber\\
&=&-i\theta(t-t')\exp\lc\int_{t'}^t\!\d t''\;
\ls-i\omega(t'')-\kappa(t'')\rs\rc,\qquad\quad\\
G^{\mathrm{A}}(t,t')&=&i\sigma\la\theta|T[\tgdd(t)\tg(t')]|\mathit{0}\ra\nonumber\\
&=&i\theta(t'-t)\exp\lc\int_t^{t'}\!\d t''\;
\ls i\omega(t'')-\kappa(t'')\rs\rc.
\eea
%
Here, $T$ is the time ordering operator.

\section{Generating functional method}
\label{GF-method}

Let us define the generating functional for the semi-free field by
\bea
Z[K,\tilde{K}]&=&
\la\theta|T\exp\lc-i\int_0^{\bar{t}}\d t\;S(t)\rc|\mathit{0}\ra,
\label{gf}
\eea
where the source function $S(t)$ reads
\bea
S(t)=\bar{K}(t)^\mu a(t)^\mu+\bar{a}(t)^\mu K(t)^\mu
=\bar{K}_\gamma(t)^\mu\gamma(t)^\mu+\bar{\gamma}(t)^\mu K_\gamma(t)^\mu,
\eea
with
%
\bea
\bar{K}(t)^\mu=\lp K(t)^*,-\tau\tilde{K}(t)\rp,\quad
K(t)^\nu=\textrm{collon}\lp K(t),\tau\tilde{K}(t)^*\rp,
\eea
%
and similar notations for $\bar{K}_\gamma(t)^\nu$ and
$K_\gamma(t)^\mu$. The $K$'s are related by the Bogoliubov transformation
\bea
\bar{K}_\gamma(t)^\nu=\bar{K}(t)^\mu[B^{-1}(t)]^{\mu\nu},\quad
K_\gamma(t)^\mu=B(t)^{\mu\nu}K(t)^\nu.
\eea
Matrix $B(t)^{\mu\nu}$ here is the one given by (\ref{tdbt}).
External fictitious fields $K(t)^\mu$, $\bar{K}(t)^\nu$ are
$c$-numbers or Grassmann numbers corresponding to $\sigma=1$ or
$\sigma=-1$, and satisfy
\bea
[K(t)^\mu,\bar{K}(t)^\nu]_{-\sigma}&=&0.
\eea
Operators $a(t)^\mu$, $\bar{a}(t)^\nu$, $\gamma(t)^\mu$
and $\bar{\gamma}(t)^\nu$ are those in the interaction
representation introduced in section~\ref{Semi-free-Ht}.

Taking the functional derivative of the generating functional
(\ref{gf}), one has
%
\bea
\delta\ln Z[K,\tilde{K}]=-i\int_0^{\bar{t}}\d t\;[\delta\bar{K}_\gamma(t)^\mu\la
\gamma(t)^\mu\ra_{_K}
+\la\bar{\gamma}(t)^\mu\ra_{_K}\delta K_\gamma(t)^\mu],
\label{func-deriv}
\eea
where $\la\gamma(t)^\mu\ra_{_K}$ and
$\la\bar{\gamma}(t)^\mu\ra_{_K}$ are defined by
%
\bea
\la\gamma(t)^\mu\ra_{_K} &=&
i\frac{\delta}{\delta\bar{K}_\gamma(t)^\mu}\ln Z[K,\tilde{K}]\nonumber\\
&=&\frac1{Z[K,\tilde{K}]}\la\theta|T[\gamma(t)^\mu\exp\{-i\int_0^{\bar{t}}\d t'
S(t')\}]|\mathit{0}\ra,\\
\la\bar{\gamma}(t)^\mu\ra_{_K}&=&
\sigma i\frac{\delta}{\delta K_\gamma(t)^\mu}\ln Z[K,\tilde{K}]\nonumber\\
&=&\frac1{Z[K,\tilde{K}]}\la\theta|T[\bar{\gamma}(t)^\mu
\exp\{-i\int_0^{\bar{t}}\d t'S(t')\}]|\mathit{0}\ra.
\eea
%
The equation of motion for $\la\gamma(t)^\mu\ra_{_K}$ is obtained
in the form
\bea
\frac{\d}{\d t}\la\gamma(t)^\mu\ra_{_K}=
[-i\omega(t)\delta^{\mu\nu}-\kappa(t)\tau_3^{\mu\nu}]\la\gamma(t)^\nu\ra_{_K}
-iK_\gamma(t)^\mu.
\eea
With the boundary conditions
%
\bea
\la\gamma(0)^{\mu=1}\ra_{_K}&=&\la\gamma(0)\ra_{_K}=0,\label{boundary-conditions0}\\
\la\gamma(\bar{t})^{\mu=2}\ra_{_K}&=&\tau\la\tgdd(\bar{t})\ra_{_K}=0,\\
\la\bar\gamma(\bar{t})^{\mu=1}\ra_{_K}&=&\la\gdd(\bar{t})\ra_{_K}=0,\\
\la\bar\gamma(0)^{\mu=2}\ra_{_K}&=&-\tau\la\tg(0)\ra_{_K}=0,
\label{boundary-conditions}
\eea
%
it can be solved as
\bea
\label{solution-a}
\la\gamma(t)^\mu\ra_{_K}&=&\int_0^{\bar{t}}\d t'\;
\mathcal{G}(t,t')^{\mu\nu}K_\gamma(t')^\nu,
\eea
where $\mathcal{G}(t,t')^{\mu\nu}$ is given by (\ref{Gmunu}). The
boundary conditions in (\ref{boundary-conditions0}) to
(\ref{boundary-conditions}) are derived by
TSCs (\ref{gamma-annihilate}) and (\ref{gamma-annihilate tilde}).

Substituting (\ref{solution-a}) into (\ref{func-deriv}), one
finally obtains \cite{Arimitsu86}
\bea
Z[K,\tilde{K}]&=&\exp\lc-i\int_0^{\bar{t}}\d t\int_0^{\bar{t}}\d t'\;
\bar{K}_\gamma(t)^\mu\mathcal{G}(t,t')^{\mu\nu}K_\gamma(t')^\nu\rc\nonumber\\
&=&\exp\lc-i\int_0^{\bar{t}}\d t\int_0^{\bar{t}}\d t'\;
\bar{K}(t)^\mu{G}(t,t')^{\mu\nu}K(t')^\nu\rc.
\eea
This expression has been derived first by Schwinger for a boson
system within the closed-time path method \cite{Schwinger61}.
Derivation of this result shown in the present section reveals the
relation between the quantum operator formalism of dissipative
fields (realized for the first time within NETFD) and their path
integral formalism \cite{Schwinger61}.

The effect of the interaction
$\hat{H}_1\equiv\hat{H}_1(a^\mu,\bar{a}^\nu)$ within the system,
which induces the dynamical correlations, can be taken into account by
the generating functional
%
%
\bea
Z_1[K,\tilde{K}]=\exp\lc-i\int_0^{\bar{t}}\d t\;\hat{H}_1
\lp i\frac{\delta}{\delta\bar{K}(t)^\mu},
\sigma i\frac{\delta}{\delta K(t)^\nu}\rp\rc Z[K,\tilde{K}].
\eea
Note that $\hat{H}_1$ should satisfy $\la\theta|\hat{H}_1 = 0$.

\section{Interaction with external fields}
\label{Interaction}

\subsection{Hermitian interaction hat-Hamiltonian}

The simplest hat-Hamiltonian representing an interaction with an external field
may be given by
\bea
\hat{H}_t'&=&H'_t-\tilde{H}'_t,\label{ief.1}
\eea
with a hermitian interaction Hamiltonian
\bea
H'_t&=&i\lp\ad b_t-b^\dagger_ta\rp,\label{ief.2}
\eea
where $b_t$, $b^\dagger_t$ and their tilde conjugates are
operators of the external system and are assumed to be
(anti-)\-com\-mu\-ta\-ti\-ve with operators $a$, $\ad$ and their tilde
conjugates of the relevant system. The subscript $t$ indicates
that these operators may depend on time. Note that the
hat-Hamiltonian (\ref{ief.1}) is tildian, i.e.
\bea
\lp i\hat{H}_t'\rp^\sim=i\hat{H}_t'.
\eea
The tilde and non-tilde operators of the external system are
related with each other by
\bea
\la|\tilde{b}^\dag_t&=&\tau^*\la|b_t,
\eea
where $\la|$ is the bra-vacuum for the external system. Applying
the bra-vacuum $\la\theta|$ for the relevant system on
(\ref{ief.1}), one has
\bea
\la\theta|\hat{H}'_t&=&-i\la\theta|
\lp\beta^{\venus}_ta+\sigma\tau\tilde{\beta}^{\venus}_t\ad\rp.
\eea
Here we introduced a new operator
\bea
\beta^{\venus}_t&=&b^\dagger_t-\tau\tilde{b}_t,
\label{beta-venus}
\eea
which annihilates the bra-vacuum $\la|$ for the external system:
\bea
\la|\beta^{\venus}_t&=&0.
\eea
As it is seen from (\ref{beta-venus}), the subscript $t$ of the
new operator $\beta^{\venus}_t$ has been inherited from the
original operators of the external system.
By applying the bra-vacuum $\la|$ on $\hat{H}'_t$ in
addition to $\la\theta|$, we observe that
\bea
\la\la\theta|\hat{H}'_t&=&0,
\eea
where the bra-vacuum of a total system is introduced by
\bea
\la\la\theta|&=&\la|\cdot\la\theta|.
\eea

The dynamics of the system is described by the Schr\"{o}\-din\-ger equation
for the ket-vacuum $|\mathit{0}(t)\ra\ra$ of the whole system:
\bea
\ddt|\mathit{0}(t)\ra\ra&=&-i\hat{H}_{t}^{\rm tot}|\mathit{0}(t)\ra\ra,
\label{Sch prime}
\eea
where $\hat{H}_{{\rm I},t}$ in $\hat{H}_{t}^{\rm tot}$ is replaced by $\hat{H}'_t$.
Conservation of the probability is guaranteed  by
$\la\la\theta| \hat{H}_{t}^{\rm tot} =0$ for the total system,
i.e. the relevant system and the external system.

\subsection{Non-Hermitian interaction hat-Hamiltonian}

Let us consider if we can have an interaction hat-Hamiltonian
which satisfies the conservation of probability within the
relevant system. This feature is consistent with the one we have 
in the case of stochastic differential equations for classical systems.

We assume that the interaction hat-Hamiltonian is
globally gauge invariant and bilinear:
\bea
\hat{H}''_t&=&i\lc h_1\ad b_t+h_2\ad\tilde{b}^\dagger_t
+h_3\ta b_t+h_4\ta\tilde{b}^\dagger_t\right.\nonumber\\
&&\left.{}+h_5\tad\tilde{b}_t+h_6\tad b^\dagger_t
+h_7a\tilde{b}_t+ h_8ab^\dagger_t\rc,\qquad\;
\eea
where quantities $h_j$ ($j=1,\cdots,8$) are time-independent
complex $c$-numbers. The tildian
\bea
\lp i\hat{H}''_t\rp^\sim&=&i\hat{H}''_t
\eea
gives us
\bea
h_1^*=h_5,\quad h_2^*=h_6,\quad h_3^*=h_7,\quad h_4^*=h_8.
\eea
By applying $\la\theta|$ from the left to the Schr\"{o}dinger equation
\bea
\ddt|\mathit{0}(t)\ra\ra&=&-i\hat{H}_t^{\rm tot} |\mathit{0}(t)\ra\ra,
\label{Sch prime prime}
\eea
with $\hat{H}_{{\rm I},t}$ in $\hat{H}_t^{\rm tot}$ being replaced by
$\hat{H}''_t$, we see that the requirement of
the conservation of probability within the relevant system leads
to
\bea
\la\theta|\hat{H}''_t&=&0.
\label{braHtpp=0}
\eea
$\hat{H}_t$ in $\hat{H}_t^{\rm tot}$ is the semi-free hat-Hamiltonian of the relevant
system satisfying (\ref{zeroev}).
From
(\ref{braHtpp=0}) we obtain
%
\bea
h_1+\sigma\tau h_3=0,&\quad&h_7+\sigma\tau h_5=0,\\
h_2+\sigma\tau h_4=0,&\quad&h_8+\sigma\tau h_6=0,
\eea
%
which are solved as
%
\bea
h_3=-\tau h_1,&\quad h_7=-\sigma\tau h_1^*,\\
h_4=-\tau h_2,&\quad h_8=-\sigma\tau h_2^*.
\eea
%
Then the structure of $\hat{H}_t''$ can be expressed in terms of
only $h_1$, $h_2$ and their complex conjugates as
\bea
\hat{H}''_t&=&i\lc\alpha^{\venus}\beta_t+
\tilde{\alpha}^{\venus}\tilde{\beta}_t\rc,
\eea
where we introduced new operators
\bea
\alpha^{\venus}&=&\ad-\tau\ta,\label{alpha-venus}\\
\beta_t&=&h_1b_t+h_2\tilde{b}^\dag_t,\label{beta1}
\eea
and their tilde conjugates. Note, that the creation operator
$\alpha^{\venus}$ annihilates the bra-vacuum $\la\theta|$:
\bea
\la\theta|\alpha^{\venus}&=&0.
\eea

In order to investigate parameters $h_1$ and $h_2$ we consider the
moments
%
\bea
\la\beta_t\tilde{\beta}_t\ra&=&(h_1+\sigma\tau h_2)
\lc\tau h_1^*\la b^\dag_tb_t\ra+h_2^*\la b_tb^\dag_t\ra\rc,\qquad\\
\la\tilde{\beta}_t\beta_t\ra&=&(h_1^*+\tau h_2^*)
\lc\sigma\tau h_1\la b^\dag_tb_t\ra+h_2\la b_tb^\dag_t\ra\rc,
\label{moments}
\eea
%
where we are using the symbol $\la\cdots\ra=\la|\cdots|t\ra$
without specifying the dynamics which determines the ket-vacuum
$|t\ra$ of the external system.  For the present purpose, the
details of its dynamics are not required.
Here we assume, however, that
the external ket-vacuum may evolve in time.
The further use of the property of the
(anti-)com\-mu\-ta\-ti\-vi\-ty, i.e.
$\la\beta_t\tilde{\beta}_t\ra=\sigma\la\tilde{\beta}_t\beta_t\ra$,
gives the necessary two relations to define $h_1$ and $h_2$:
%
\bea
(\tau h_1+h_2)h_1^*&=&(\tau h_1^*+\sigma h_2^*)h_1,\\
\sigma(\tau h_1+h_2)h_2^*&=&(\tau h_1^*+\sigma h_2^*)h_2,
\eea
%
which reduce to
\bea
h_1^*h_2&=&\sigma h_1h_2^*.
\label{requirement}
\eea
We can express $h_1$ and $h_2$ as
\bea
\label{h1-h2}
h_1=\mu\e^{i\theta_1},\quad
h_2=\nu\e^{i\theta_2},
\eea
where $\mu,\nu\in\mathbf{R}$, namely $\mu=|h_1|$, $\nu=|h_2|$.
From the requirement (\ref{requirement}), one has
$\theta_2=\theta_1$ for $\sigma=1$ and $\theta_2=\theta_1-\pi/2$
for $\sigma=-1$. 
Substituting (\ref{h1-h2}) into (\ref{beta1}) and putting the
phase factor $\e^{i\theta_1}$ into $b_t$ and $\tilde{b}^\dag_t$, we
have
\bea
\beta_t&=&\mu b_t+\sigma\tau\nu\tilde{b}^\dag_t.
\eea
Thus, the vector $\la|\beta_t$ is calculated as
\bea
\la|\beta_t&=&\la| (\mu b_t+
\sigma\tau\nu \tilde{b}^\dag_t )
=(\mu+\sigma\nu)\la|b_t.
\eea
The further requirement that the
norm of $\la|\beta_t$ should be equal to that of $\la|b_t$, i.e.
$\|\la|\beta_t\|=\|\la|b_t\|$, leads one to the relation
\bea
\mu+\sigma\nu&=&1.\label{ief.26}
\eea

\subsection{Relation between the two interaction hat-Hamiltonians}

Note that the hermitian interaction hat-Hamiltonian $\hat{H}_t'$
and the non-hermitian one $\hat{H}_t''$ are related to each other
by
\bea
\hat{H}'_t=\hat{H}''_t-i\lc\beta^{\venus}_t(\mu a +\sigma\tau\nu\tad)
+\tilde{\beta}^{\venus}_t(\mu\ta+\sigma\tau^*\nu\ad)\rc.
\label{ief.28}
\eea

With an auxiliary parameter $0\leq \lambda\leq 1$, it is
possible to make a simultaneous consideration of both hermitian
and non-hermitian interaction hat-Hamiltonians by introducing
\bea
\hat{H}_{\mathrm{I},t}=i\lc\alpha^{\venus}\beta_t
+\tilde{\alpha}^{\venus}\tilde{\beta}_t\rc
-i\lambda\lc\beta^{\venus}_t(\mu a+\sigma\tau\nu\tad)
+\tilde{\beta}^{\venus}_t(\mu\ta+\sigma\tau^*\nu\ad)\rc.
\qquad\label{H-lambda}
\eea
It is easy to see that this expression is reduced to $\hat{H}_t'$
in the case $\lambda=1$ and to $\hat{H}_t''$ in the case
$\lambda=0$, respectively.
The dynamics of the system is now described by the Shr\"odinger equation 
(\ref{Sch prime}) or (\ref{Sch prime prime}) with $\hat{H}_{{\rm I},t}$ 
in $\hat{H}_t^{\rm tot}$ being given by (\ref{H-lambda}).

\section{Quantum stochastic differential equations}
\label{QSDE}

\subsection{Quantum stochastic Liouville equations}

\subsubsection{It\^o type}

Let us derive the general form of the semi-free hat-Hamiltonian
$\hat{\mathcal{H}}_{F,t}\d t$ for a stochastic Liouville equation of
the It\^o type
\bea
\d|\mathit{0}_F(t)\ra&=&-i\hat{\mathcal{H}}_{F,t}\d t
|\mathit{0}_F(t)\ra \label{sde.1}
\eea
where a subscript $F$ is added to indicate that we are considering a system 
under the influence of a random force. 
We assume that the hat-Hamiltonian $\hat{\mathcal{H}}_{F,t}\d t$
for the stochastic semi-free
field is bilinear in $a$, $\ad$, $\d F_t$, $\d F^\dag_t$ and their
tilde conjugates, and that it is invariant under the phase transformation
$a\rightarrow a\e^{i\phi}$ and $\d F_t\rightarrow\d
F_t\e^{i\phi}$. Here, $a$, $\ad$ and their tilde conjugates are
operators of a relevant system satisfying the canonical
(anti-)\-com\-mu\-ta\-ti\-on relation
\bea
[a,\ad]_{-\sigma}&=&1,
\eea
whereas $\d F_t$, $\d F^\dag_t$ and their tilde conjugates are
random force operators. The tilde and non-tilde operators are
related with each other by the TSC
%
\bea
\la\theta|\tad&=&\tau^*\la\theta|a,\\
\la|\d\tilde{F}^\dag_t&=&\tau^*\la|\d F_t,\label{TSC-Ft}
\eea
%
where $\la\theta|$ and $\la|$ are, respectively, the thermal
bra-vacuum of the relevant system and of the random force.

From the investigation in section \ref{Interaction}, we can propose that the
required form of the hat-Ha\-mil\-to\-ni\-an should be
\bea
\hat{\mathcal{H}}_{F,t}\d t&=&\hat{H}_t\d t+\d\hat{M}_t,
\eea
where $\hat{H}_t$ is specified by (\ref{Hs_Pi}) with
$\hat{\mPi}_t$ having the same structure as (\ref{Pit2}) or
(\ref{Pit3}). 
For later convenience, we rewrite $\hat{\mPi}_t$ as
\bea
\hat{\mPi}_t&=&\hat{\mPi}_{\mathrm{R}}+\hat{\mPi}_{\mathrm{D}}
\eea
with
%
\bea
\hat{\mPi}_{\mathrm{R}}&=&-\kappa(t)\lc\alpha^{\venus}\alpha+
\tilde{\alpha}^{\venus}\tilde{\alpha}\rc,\\
\hat{\mPi}_{\mathrm{D}}&=&{}\sigma\tau\lc2\kappa(t)[n(t)+\eta]+\dot{n}(t)\rc
\alpha^{\venus}\tilde{\alpha}^{\venus},
\eea
%
where we introduced
\bea
\alpha&=&\xi a+\sigma\tau\eta\tilde{a}^\dag,
\qquad\xi+\sigma\eta=1,
\label{alpha}
\eea
which forms a canonical set with $\alpha^{\venus}$ defined by
(\ref{alpha-venus}), i.e.
\bea
[\alpha,\alpha^{\venus}]_{-\sigma}&=&1.
\label{alpha-alpha-venus-sigma-commutator}
\eea
The one-particle distribution function $n(t)$ is defined by
\bea
n(t)=\Big\la\la\theta|a^\dag a|\mathit{0}_F(t)\ra\Big\ra,
\eea
and satisfies the Boltzmann equation (\ref{crf.6}) (see Appendix \ref{crf}).
Here, $\la\cdots\ra$ means to take the random average, i.e., the vacuum
expectation value with respect to the thermal bra- and ket-vacuums
of random force: $\la\cdots\ra=\la|\cdots|\ra$. Terms
$\hat{\mPi}_{\mathrm{R}}$ and $\hat{\mPi}_{\mathrm{D}}$ are,
respectively, the relaxational and diffusive parts of the damping
operator $\hat{\mPi}_t$. 

The martingale $\d\hat{M}_t$ is the term
containing operators representing quantum Brownian motion and
satisfies
\bea
\la\d\hat{M}_t\ra&=&0.
\eea
Associating $\d F_t$ and $\d F_t^\dag$ with $b_t\d t$ and $b_t^\dag\d t$
in (\ref{H-lambda}), respectively, we have
\bea
\d\hat{M}_t&=&i\lc\alpha^{\venus}\d W_t
+\tilde{\alpha}^{\venus}\d\tilde{W}_t\rc\nonumber\\
&&-i\lambda\lc\d W_t^{\venus}(\mu a+\sigma\tau\nu\tad)
+\d\tilde{W}_t^{\venus}(\mu\ta+\sigma\tau^*\nu\ad)\rc,
\eea
where we introduced new operators
%
\bea
\d W_t&=&\mu\d F_t+\sigma\tau\nu\d\tilde{F}^\dag_t,\label{sde.11}\\
\d W_t^{\venus}&=&\d F_t^\dag-\tau\d\tilde{F}_t.\label{Wt-venus}
\label{dWt}
\eea
%
Note that $\d W^{\venus}_t$ and $\d\tilde{W}^{\venus}_t$
annihilate the bra-vacuum for random force $\la|$:
\bea
\la|\d W^{\venus}_t&=&0.\label{bra-w-venus-zero}
\eea

In the It\^o multiplication, the random force operators $\d W_t$,
$\d W_t^{\venus}$ and their tilde conjugates do not correlate with
quantities at time $t$, e.g. $|\mathit{0}_F(t)\ra$:
\bea
\La\d\hat{M}_t|\mathit{0}_F(t)\ra\Ra=0.
\eea
Thus, taking the random average of the stochastic Liouville
equation (\ref{sde.1}), we arrive at the Fokker-Planck equation
\bea
\ddt|\mathit{0}(t)\ra&=&-i\hat{H}_t|\mathit{0}(t)\ra,\label{sde.15}
\eea
where $|\mathit{0}(t)\ra=\Big\la|\mathit{0}_F(t)\ra\Big\ra$.

The formal solution of (\ref{sde.1}) can be written as
\bea
|\mathit{0}_F(t)\ra&=&\hat{V}_F(t)|\mathit{0}_F(0)\ra,
\label{formal-solution}
\eea
where the time-evolution generator is defined through
\bea
\d\hat{V}_F(t)&=&-i\hat{\mathcal{H}}_{F,t}\d t\hat{V}_F(t)\label{sde.20}
\eea
with the initial condition $\hat{V}_F(0)=1$.

\subsubsection{Fluctuation-dissipation theorem of the second kind}

By making use of the relation between the It\^o and the
Stratonovich stochastic multiplications (see
Ap\-pen\-dix~\ref{IS}), we can rewrite the It\^o type stochastic
Liouville equation into the Stratonovich type as follows. Relation
(\ref{IS.4b}) makes the term containing random force operators in
the r.h.s of (\ref{sde.1}) be
\bea
\d\hat{M}_t|\mathit{0}_F(t)\ra&=&\d\hat{M}_t\circ|\mathit{0}_F(t)\ra
-\frac12\d\hat{M}_t\,\d|\mathit{0}_F(t)\ra,\qquad
\label{connections}
\eea
where the symbol $\circ$ has been introduced to indicate the
Stratonovich stochastic multiplication (see
Appendix~\ref{IS}). Substituting (\ref{sde.1}) into the last term
for $\d|\mathit{0}_F(t)\ra$ and neglecting terms of the higher
order than $\d t$, we arrive at the quantum stochastic Liouville
equation of the Stratonovich type 
\bea
\d|\mathit{0}_F(t)\ra&=&-i\hat{H}_{F,t}\d t\circ|\mathit{0}_F(t)\ra,
\label{sde.18}
\eea
with
%
\begin{eqnarray}
\hat{H}_{F,t}\d t&=&\hat{H}_{\mathrm{S},t}\d t
+i\hat{\mPi}_t\d t+\d\hat{M}_t+\frac{i}2\d\hat{M}_t\d\hat{M}_t
\qquad\label{H-St-a}\\
&=&\hat{H}_{\mathrm{S},t}\d t+i(1-\lambda)\hat{\mPi}_{\mathrm{R}}\d t
+\d\hat{M}_t.\label{H-St-b}
\end{eqnarray}
%
In order to obtain expression (\ref{H-St-b}) we used
the generalized fluctuation-dissipation
theorem of the second kind, which can be written as
\bea
\d\hat{M}_t\d\hat{M}_t&=&-2\lp\lambda\hat{\mPi}_{\mathrm{R}}
+\hat{\mPi}_{\mathrm{D}}\rp\d t,\label{GFDT}
\eea
(refer to Appendix \ref{crf} for derivation).
Note that in the Stratonovich multiplication random force operators
$\d W_t$, $\d W_t^{\venus}$ and their tilde conjugates correlate
with quantities at time $t$, i.e.,
\bea
\La\d\hat{M}_t\circ|\mathit{0}_F(t)\ra\Ra\ne0.
\eea

The formal solution of (\ref{sde.18}) has the form
(\ref{formal-solution}), where the stochastic time-evolution
generator $\hat{V}_F(t)$ is defined through
\bea
\d\hat{V}_F(t)&=&-i\hat{H}_{F,t}\d t\circ\hat{V}_F(t)\label{Vhat-S}
\eea
with the initial condition $\hat{V}_F(0)=1$.

\subsubsection{Correlations of the random force operators}

Operators $\d W_t$, $\d W_t^{\venus}$ and their tilde conjugates are of the
quantum stochastic Wiener process satisfying (for derivation of the
results see Appendix \ref{crf})
%
\bea
\la\d W_t\ra&=&\la\d\tilde{W}_t\ra=0,\\
\la\d W_t\d W_s\ra&=&\la\d\tilde{W}_t\d\tilde{W}_s\ra=0,\label{dwdtw0}\\
\la\d W_t\d\tilde{W}_s\ra&=&\sigma\la\d\tilde{W}_s\d W_t\ra\nonumber\\
&=&\tau\lc2\kappa(t)[n(t)+\nu]+\dot{n}(t)\rc\delta(t-s)\d t\;\d s,
\label{dwdtw}
\eea
%
and
%
\bea
\la\d W^{\venus}_t\ra&=&\la\d\tilde{W}^{\venus}_t\ra=0,\\
\la\d W^{\venus}_t\d W^{\venus}_s\ra&=&\la\d W^{\venus}_t\d\tilde{W}^{\venus}_s\ra=0,
\label{dwdtwd0}\\
\la\d W^{\venus}_t\d W_s\ra&=&\la\d\tilde{W}^{\venus}_t\d\tilde{W}_s\ra=0,\\
\la\d W_t\d W^{\venus}_s\ra&=&\la\d\tilde{W}_t\d\tilde{W}^{\venus}_s\ra
=2\kappa(t)\delta(t-s)\d t\;\d s.\qquad\quad
\label{dwdtwd}
\eea
%
Due to the argument of Appendix \ref{crf} we also have $\xi=\mu$
and $\eta=\nu$, which leads to
\bea
\alpha=\mu a+\sigma\tau\nu\tad.
\eea
In the following, we will use this definition in both
$\hat{\mPi}_t$ and $\d\hat{M}_t$. Especially, the latter becomes
\bea
\d\hat{M}_t=i\lc\alpha^{\venus}\d W_t
+\tilde{\alpha}^{\venus}\d\tilde{W}_t\rc
-i\lambda\lc\d W_t^{\venus}\alpha
+\d\tilde{W}_t^{\venus}\tilde{\alpha}\rc.\label{dM def}
\eea
It is important to note here that the martingale $\d\hat{M}_t$ is introduced 
in the normal ordering with respect to all operators
$\alpha^{\venus}$, $\alpha$, $dW_t^{\venus}$, $dW_t$ and their tilde 
conjugates.

Within the weak relations, the correlations (\ref{dwdtw0}), (\ref{dwdtw}) and
(\ref{dwdtwd0}) to (\ref{dwdtwd}) reduce, respectively, to
%
\bea
\d W_t\d W_s&=&\d\tilde{W}_t\d\tilde{W}_s=0,\label{sde.140}\\
\d W_t\d\tilde{W}_s&=&\sigma\d\tilde{W}_s\d W_t
=\tau\lc2\kappa(t)[n(t)+\nu]+\dot{n}(t)\rc\delta(t-s)\d s\;\d t,
\label{sde.14}
\eea
%
and
%
\bea
\d W^{\venus}_t\d W^{\venus}_s&=&\d W^{\venus}_t\d\tilde{W}^{\venus}_s=0,\\
\d W^{\venus}_t\d W_s&=&\d\tilde{W}^{\venus}_t\d\tilde{W}_s=0,\\
\d W_t\d W^{\venus}_s&=&\d\tilde{W}_t\d\tilde{W}^{\venus}_s
=2\kappa(t)\delta(t-s)\d s\;\d t.\qquad
\label{utg.8}
\eea
%

\subsection{Stochastic semi-free operators}

The stochastic semi-free operators are defined by
\bea
A(t)&=&\hat{V}_F^{-1}(t)A\hat{V}_F(t),\label{A(t)}
\eea
whereas the random force operators in the Heisenberg
representation by
\bea
\label{W(t)a}
W(t)=\hat{V}_F^{-1}(t)W_t\hat{V}_F(t),\quad
W^{\venus}(t)=\hat{V}_F^{-1}(t)W_t^{\venus}\hat{V}_F(t),
\eea
and their tilde conjugates.
We also use the convenient operators introduced by
\bea
\label{W(t)b}
\dbar W(t)=\hat{V}_F^{-1}(t)\d W_t\hat{V}_F(t),\;
\dbar W^{\venus}(t)=\hat{V}_F^{-1}(t)\d W_t^{\venus}\hat{V}_F(t),
\eea
and their tilde conjugates. Here,
\bea
\d\hat{V}_F^{-1}(t)&=&i\hat{V}_F^{-1}(t)\hat{\mathcal{H}}_{F,t}^-\d t,\label{sde.20m}
\eea
with $\hat{V}_F^{-1}(0)=1$. $\hat{\mathcal{H}}_{F,t}^-\d t$ is
specified by
\bea
\hat{\mathcal{H}}_{F,t}^-\d t=\ds\hat{\mathcal{H}}_{F,t}\d t
+i\d\hat{M}_t\d\hat{M}_t
=\hat{H}_{\mathrm{S},t}\d t+i\hat{\mPi}_t^-\d t+\d\hat{M}_t
\eea
with
\bea
\hat{\mPi}_t^-&=&(1-2\lambda)\hat{\mPi}_{\mathrm{R}}
-\hat{\mPi}_{\mathrm{D}}.
\eea

In particular cases when $A$ represents $a$ or $a^\dag$ we have
\bea
a(t)=\hat{V}_F^{-1}(t)a\hat{V}_F(t),\quad
\tadd(t)=\hat{V}_F^{-1}(t)\tad\hat{V}_F(t).
\eea
Since the stochastic tildian hat-Hamiltonian
$\hat{\mathcal{H}}_{F,t}\d t$ is not necessarily hermitian, we
introduced the symbol $\dag\!\dag$ in order to distinguish it from
the hermite conjugation $\dag$. It is as\-su\-med that, at initial
time $t=0$, the relevant system starts to contact with the
irrelevant system representing the stochastic process described by
the random force operators $\d F_t$, $\d F_t^\dag$ and their tilde
conjugates. Within the formalism, the random force operators $\d
F_t$ and $\d F^\dag_t$ are assumed to (anti-)commute with any
relevant system operator $A$ in the Schr\"o\-din\-ger
representation, i.e.
\bea
[A,\d F_t\}=0,\quad[A,\d F^\dag_t\}=0.
\eea
The semi-free operators $a(t)$, $\add(t)$ and their tilde
conjugates keep the equal-time canonical (anti-)commutation
relations
\bea
[a(t),\add(t)]_{-\sigma}&=&1,
\eea
and satisfy TSC
\bea
\la\la\theta|\tadd(t)&=&\tau^*\la\la\theta|a(t).
\eea

Calculating the time derivatives of Heisenberg operators of 
the quantum Brownian motion
(\ref{W(t)a}) within the It\^o calculus (\ref{IS.6}), 
and taking into account (\ref{sde.20}),
(\ref{sde.20m}) with the characteristics of the It\^o multiplication
%
\bea
[\d W_t^{\venus}, W_t]_{-\sigma}=[\d W_t, W_t^{\venus}]_{-\sigma}&=&0,\label{hoqbm.1a}\\\relax
[\d W_t,\hat{V}_F(t)]=[\d W_t^{\venus},\hat{V}_F(t)]&=&0,\label{hoqbm.1b}
\label{hoqbm.1}
\eea
%
and their tilde conjugates, one has
%
\bea
\d W(t)&=&\d W_t-i\hat{V}_F^{-1}(t)[\d W_t,\d\hat{M}_t]\hat{V}_F(t),\label{hoqbm.2a}\\
\d W^{\venus}(t)&=&\d W_t^{\venus}-i\hat{V}_F^{-1}(t)[\d W_t^{\venus},\d\hat{M}_t]
\hat{V}_F(t),\qquad\label{hoqbm.2b}
\label{hoqbm.2}
\eea
%
and their tilde conjugates. With the help of (\ref{sde.140}) to
(\ref{utg.8}), the expressions (\ref{hoqbm.2a}) and (\ref{hoqbm.2}) reduce, respectively, to
%
\bea
\d W(t)&=&\d W_t-2\lambda\kappa(t)\alpha(t)\d t,
\label{dW(t)0}\\
\d W^{\venus}(t)&=&\d W^{\venus}_t-2\kappa(t)\alpha^{\venus}(t)\d t,
\label{dW(t)}
\eea
%
(and their tilde conjugates),
while (\ref{hoqbm.1b}) gives
\bea
\label{dbarW(t)}
\dbar W(t)=\d W_t,\quad
\dbar W^{\venus}(t)=\d W_t^{\venus},
\eea
(and their tilde conjugates). 

We see that the martingale operator
$\dbar\hat{M}(t)\equiv\hat{V}_F^{-1}(t)\d\hat{M}_t\hat{V}_F(t)$
being written in terms of the Heisenberg operators reads
\bea
\dbar\hat{M}(t)&=&i\lc\alpha^{\venus}(t)\dbar W(t)
+\tilde{\alpha}^{\venus}(t)\dbar\tilde{W}(t)\rc\nonumber\\
&&{}-i\lambda\lc\dbar W^{\venus}(t)\alpha(t)
+\dbar\tilde{W}^{\venus}(t)\tilde{\alpha}(t)\rc\nonumber\\
&=&i\lc\alpha^{\venus}(t)\d W(t)
+\tilde{\alpha}^{\venus}(t)\d\tilde{W}(t)\rc\nonumber\\
&&{}-i\lambda\lc\d W^{\venus}(t)\alpha(t)
+\d\tilde{W}^{\venus}(t)\tilde{\alpha}(t)\rc\nonumber\\
&\equiv&\d\hat{M}(t),\label{property}
\eea
and keeps the property
\bea
\la\dbar\hat{M}(t)\ra=\la\d\hat{M}(t)\ra&=&0
\eea
for arbitrary $\lambda$. 
The beautiful relation (\ref{property}) is manifestations of 
the normal ordered definition (\ref{dM def}).
Note that the increments in the martingale (\ref{dM def}) are 
introduced just through the random force operators
$dW_t^{\venus}$, $dW_t$ and their tilde conjugates.
Therefore, $d\hat{M}(t)$ is different from the operator calculated
by $\d\lp i\lc\alpha^{\venus}(t) W(t)
+\mbox{t.c.}\rc 
-i\lambda\lc W^{\venus}(t)\alpha(t)
+ \mbox{t.c.}\rc\rp$. 
Here, t.c.\ indicates the tilde conjugation.

\subsection{Quantum Langevin equations}

\subsubsection{It\^o type}

Substituting $\d\hat{V}_F(t)$, (\ref{sde.20}),
and $\d\hat{V}^{-1}_F(t)$, (\ref{sde.20m}),
into the time derivative of the dynamical
quantity $A(t)$ within the It\^o calculus
(\ref{IS.6}), we obtain the quantum Langevin equation of the It\^o type 
in the form
%
\bea
\d A(t)&=&i[\hat{\mathcal{H}}_F(t)\d t,A(t)]
-\d\hat{M}(t)[\d\hat{M}(t),A(t)]\label{sde.38}\\
&=&i[\hat{H}_{\mathrm{S}}(t),A(t)]\d t\nonumber\\
&&+\kappa(t)\lp\alpha^{\venus}(t)[\alpha(t),A(t)\}
+\tilde{\alpha}^{\venus}(t)[\tilde{\alpha}(t),A(t)\}\right.\nonumber\\
&&+\left.(2\lambda-1)\big([A(t),\alpha^{\venus}(t)\}\alpha(t)
+[A(t),\tilde{\alpha}^{\venus}(t)\}\tilde{\alpha}(t)\big)\rp\d t\nonumber\\
&&+\tau\big(2\kappa(t)[n(t)+\nu]+\dot{n}(t)\big)[\tilde{\alpha}^{\venus}(t),
[\alpha^{\venus}(t),A(t)\}\}\d t\nonumber\\
&&+[A(t),\alpha^{\venus}(t)\}\d W(t)
+[A(t),\tilde{\alpha}^{\venus}(t)\}\d\tilde{W}(t)\nonumber\\
&&+\lambda\lp\d W^{\venus}(t)[\alpha(t),A(t)\}
+\d\tilde{W}^{\venus}(t)[\tilde{\alpha}(t),A(t)\}\rp
\label{out}\\
&=&i[\hat{H}_{\mathrm{S}}(t),A(t)]\d t\nonumber\\
&&-\kappa(t)\lp[A(t),\alpha^{\venus}(t)\}\alpha(t)
+[A(t),\tilde{\alpha}^{\venus}(t)\}\tilde{\alpha}(t)\right.\nonumber\\
&&+\left.(2\lambda-1)\big(\alpha^{\venus}(t)[\alpha(t),A(t)\}
+\tilde{\alpha}^{\venus}(t)[\tilde{\alpha}(t),A(t)\}\big)\rp\d t\nonumber\\
&&+\tau\big(2\kappa(t)[n(t)+\nu]+\dot{n}(t)\big)[\tilde{\alpha}^{\venus}(t),
[\alpha^{\venus}(t),A(t)\}\}\d t\nonumber\\
&&+[A(t),\alpha^{\venus}(t)\}\d W_t
+[A(t),\tilde{\alpha}^{\venus}(t)\}\d\tilde{W}_t\nonumber\\
&&+\lambda\lp\d W^{\venus}_t[\alpha(t),A(t)\}
+\d\tilde{W}^{\venus}_t[\tilde{\alpha}(t),A(t)\}\rp,
\label{in}
\eea
%
where (\ref{sde.140}) to (\ref{utg.8}) for
multiplications among the random force operators are employed. To derive
(\ref{in}) from (\ref{out}), we used (\ref{dW(t)0}) and (\ref{dW(t)}). Note that the
Langevin equations (\ref{out}) and (\ref{in}) written, respectively, by means of
the quantum Brownian motion in the Heisenberg representation and
by means of that in the Schr\"odinger representation 
may be related with the ``out'' and ``in'' fields introduced by
Gardiner~\textit{et al.}~\cite{Gardiner85,Gardiner00}.

With the help of (\ref{out}) one can verify that the calculus
rule for the product of arbitrary relevant stochastic operators,
say $A(t)$ and $B(t)$, satisfies the It\^o calculus (\ref{IS.6}).
This proves that QSDE
(\ref{out}) is of the It\^o type indeed. Furthermore, since 
(\ref{out}) is the time-evolution equation for any relevant stochastic
operator $A(t)$, it is It\^o's formula for quantum systems.

\subsubsection{Stratonovich type}

The quantum Langevin equation of the Stratonovich type can be derived
similarly if one starts from the expression for a dynamical
quantity $A(t)$, (\ref{A(t)}), and considers its derivative in the
Stratonovich calculus (\ref{IS.7}) with the help of
\bea
\d\hat{V}_F^{-1}(t)&=&i\hat{V}_F^{-1}(t)\circ\hat{H}_{F,t}\d t.\label{Vhat-minus-S}
\eea
Substituting (\ref{Vhat-S}) for $\d\hat{V}_F(t)$ and
(\ref{Vhat-minus-S}) for $\d\hat{V}^{-1}_F(t)$ into $\d A(t)$, we have for 
the stochastic Heisenberg equation of the Stratonovich type
%
\bea
\d A(t)&=&i[\hat{H}_F(t)\d t\stackrel{\circ}{,}A(t)]\label{sde.29}\\
&=&i[\hat{H}_{\mathrm{S}}(t),A(t)]\d t\nonumber\\
&&+\kappa(t)\lp\alpha^{\venus}(t)[\alpha(t),A(t)\}
+\tilde{\alpha}^{\venus}(t)[\tilde{\alpha}(t),A(t)\}\right.\nonumber\\
&&+\left.(2\lambda-1)\big([A(t),\alpha^{\venus}(t)\}\alpha(t)
+[A(t),\tilde{\alpha}^{\venus}(t)\}\tilde{\alpha}(t)\big)\rp\d t\nonumber\\
&&+[A(t),\alpha^{\venus}(t)\}\circ\d W(t)
+[A(t),\tilde{\alpha}^{\venus}(t)\}\circ\d\tilde{W}(t)\nonumber\\
&&+\lambda\lp\d W^{\venus}(t)\circ[\alpha(t),A(t)\}
+\d\tilde{W}^{\venus}(t)\circ[\tilde{\alpha}(t),A(t)\}\rp
\label{S-out}\\
&=&i[\hat{H}_{\mathrm{S}}(t),A(t)]\d t\nonumber\\
&&-\kappa(t)\lp[A(t),\alpha^{\venus}(t)\}\alpha(t)
+[A(t),\tilde{\alpha}^{\venus}(t)\}\tilde{\alpha}(t)\right.\nonumber\\
&&+\left.(2\lambda-1)\big(\alpha^{\venus}(t)[\alpha(t),A(t)\}
+\tilde{\alpha}^{\venus}(t)[\tilde{\alpha}(t),A(t)\}\big)\rp\d t\nonumber\\
&&+[A(t),\alpha^{\venus}(t)\}\circ\d W_t
+[A(t),\tilde{\alpha}^{\venus}(t)\}\circ\d\tilde{W}_t\nonumber\\
&&+\lambda\lp\d W^{\venus}_t\circ[\alpha(t),A(t)\}
+\d\tilde{W}^{\venus}_t\circ[\tilde{\alpha}(t),A(t)\}\rp.
\label{S-in}
\eea
%
Here, we defined
\bea
[X(t)\stackrel{\circ}{,}Y(t)]&=&X(t)\circ Y(t)-Y(t)\circ X(t)
\eea
for arbitrary operators $X(t)$ and $Y(t)$, and
\bea
\hat{H}_F(t)\d t&=&\hat{V}_F^{-1}(t)\circ\hat{H}_{F,t}\d t\circ\hat{V}_F(t).
\eea
Note that
\bea
\hat{V}_F^{-1}(t)\circ\d\hat{M}_t\circ\hat{V}_F(t)&=&
\hat{V}_F^{-1}(t)\d\hat{M}_t\hat{V}_F(t)\nonumber\\
&&+\frac12\hat{V}_F^{-1}(t)\d\hat{M}_t\d\hat{V}_F(t)
+\frac12\d\hat{V}_F^{-1}(t)\d\hat{M}_t\hat{V}_F(t)\nonumber\\
&=&\d\hat{M}(t).
\eea
Using expression (\ref{S-out}), one can readily verify that the
calculus rule for the product of arbitrary relevant system
operators, say $A(t)$ and $B(t)$, satisfies the Stratonovich type
calculus (\ref{IS.7}). This fact proves that QSDE 
(\ref{S-out}) is indeed of the
Stratonovich type, and provides us with the reason
why the stochastic Heisenberg equation 
(\ref{sde.29}) has the same structure as the one (\ref{HEDS}) 
for non-stochastic operators.

The quantum Langevin equation of the Stratonovich type can be also
derived from that of the It\^o type by making use the
connection formulae (\ref{IS.3a}) and (\ref{IS.3b}). When $\d Y(t)$ is $\d W(t)$ or
$\d\tilde{W}(t)$, and $X(t)$ is constituted by the relevant
operator, say $A(t)$, satisfying the quantum Langevin equation of
the It\^o type, the connection formula (\ref{IS.3a}) reduces,
respectively, to
%
\bea
A(t)\cdot\d W(t)&=&A(t)\circ\d W(t)\nonumber\\
&&-\frac12\sigma\tau\big(2\kappa(t)[n(t)+\nu]+\dot{n}(t)\big)
[A(t),\tilde{\alpha}^{\venus}(t)\}\d t,
\label{s-to-ito}\\
A(t)\cdot\d\tilde{W}(t)&=&A(t)\circ\d\tilde{W}(t)\nonumber\\
&&-\frac12\tau\big(2\kappa(t)[n(t)+\nu]+\dot{n}(t)\big)
[A(t),\alpha^{\venus}(t)\}\d t.\label{s-to-ito2}
\eea
%
Similarly, when $\d X(t)$ is $\d W^{\venus}(t)$ or
$\d\tilde{W}^{\venus}(t)$, and $Y(t)$ is $A(t)$, the connection
formula (\ref{IS.3b}) reduces, respectively, to
%
\bea
\d W^{\venus}(t)\cdot A(t)&=&\d W^{\venus}(t)\circ A(t),\label{s-to-itov1}\\
\d\tilde{W}^{\venus}(t)\cdot A(t)&=&\d\tilde{W}^{\venus}(t)\circ A(t).\label{s-to-itov2}
\eea
%
Using these relations in (\ref{out}), the quantum Langevin
equation of the Stratonovich type is obtained in the form
(\ref{S-out}).

Substituting $\alpha$ and $\tilde{\alpha}^{\venus}$ for $A$ as an
example, we see that both (\ref{sde.38}) and (\ref{sde.29}) result
in
%
\bea
\d\alpha(t)&=&-i\omega(t)\alpha(t)\d t
-(1-2\lambda)\kappa(t)\alpha(t)\d t+\d W(t)\nonumber\\
&=&-\ls i\omega(t)+\kappa(t)\rs\alpha(t)\d t+\d W_t,\\
\d\tilde{\alpha}^{\venus}(t)&=&
-i\omega(t)\tilde{\alpha}^{\venus}(t)\d t
+\kappa(t)\tilde{\alpha}^{\venus}(t)\d t
+\lambda\d\tilde{W}^{\venus}(t)\nonumber\\
&=&-\ls i\omega(t)-(1-2\lambda)\kappa(t)\rs\tilde{\alpha}^{\venus}(t)\d t
+\lambda\d\tilde{W}_t^{\venus},
\eea
%
which are written in terms of the original operators as
%
\bea
\d a(t)&=&
-\ls i\omega(t)-\lambda\kappa(t)\rs a(t)\d t\nonumber\\
&&{}-(1-\lambda)\kappa(t)[(\mu-\sigma\nu)a(t)+2\sigma\tau\nu\tadd(t)]\d t
\nonumber\\
&&{}+\d W(t)-\lambda\sigma\tau\nu\d\tilde{W}^{\venus}(t)\nonumber\\
&=&-\ls i\omega(t)+\lambda\kappa(t)\rs a(t)\d t\nonumber\\
&&{}-(1-\lambda)\kappa(t)[(\mu-\sigma\nu)a(t)+2\sigma\tau\nu\tadd(t)]\d t
\nonumber\\
&&{}+\d W_t-\lambda\sigma\tau\nu\d\tilde{W}_t^{\venus},\\
\d\tadd(t)&=&-\ls i\omega(t)-\lambda\kappa(t)\rs\tadd(t)\d t\nonumber\\
&&{}+(1-\lambda)\kappa(t)[(\mu-\sigma\nu)\tadd(t)-2\sigma\tau\mu a(t)]\d t
\nonumber\\
&&{}+\sigma\tau\d W(t)+\lambda\mu\d\tilde{W}^{\venus}(t)\nonumber\\
&=&-\ls i\omega(t)+\lambda\kappa(t)\rs\tadd(t)\d t\nonumber\\
&&{}+(1-\lambda)\kappa(t)[(\mu-\sigma\nu)\tadd(t)-2\sigma\tau\mu a(t)]\d t
\nonumber\\
&&{}+\sigma\tau\d W_t+\lambda\mu\d\tilde{W}_t^{\venus}.
\label{hem2}
\eea
%
These equations are the same in both It\^o and Stra\-to\-no\-vich
multiplications as they should be with the martingale (\ref{dM def}).

\subsection{Averaged equation of motion}

Applying the total bra-vacuum $\la\la\theta|$ to the It\^o type
quantum Lan\-ge\-vin equ\-a\-ti\-on (\ref{sde.38}) to (\ref{in}), 
one can derive
the stochastic equation of motion of the It\^o type for the
bra-vector state $\la\la\theta|A(t)$ in the form
\bea
\d\la\la\theta|A(t)&=&i\la\la\theta|[\hat{H}_{\mathrm{S}}(t),A(t)]\d t\nonumber\\
&&+(2\lambda-1)\kappa(t)\la\la\theta|\lp A(t)[\alpha^{\venus}(t)\alpha(t)+
\tilde{\alpha}^{\venus}(t)\tilde{\alpha}(t)]\rp\d t\nonumber\\
&&+\sigma\tau\big(2\kappa(t)[n(t)+\nu]+\dot{n}(t)\big)\la\la\theta|
A(t)\alpha^{\venus}(t)\tilde{\alpha}^{\venus}(t)\d t\nonumber\\
&&+\la\la\theta|\lp A(t)[\alpha^{\venus}(t)\d W(t)
+\tilde{\alpha}^{\venus}(t)\d\tilde{W}(t)]\rp\\
&=&i\la\la\theta|[\hat{H}_{\mathrm{S}}(t),A(t)]\d t\nonumber\\
&&-\kappa(t)\la\la\theta|\lp A(t)[\alpha^{\venus}(t)\alpha(t)+
\tilde{\alpha}^{\venus}(t)\tilde{\alpha}(t)]\rp\d t\nonumber\\
&&+\sigma\tau\big(2\kappa(t)[n(t)+\nu]+\dot{n}(t)\big)\la\la\theta|
A(t)\alpha^{\venus}(t)\tilde{\alpha}^{\venus}(t)\d t\nonumber\\
&&+\la\la\theta|\lp A(t)[\alpha^{\venus}(t)\d W_t
+\tilde{\alpha}^{\venus}(t)\d\tilde{W}_t]\rp.
\label{aem-ito}
\eea
In terms of operators $a(t)$ and $\add(t)$ it becomes
\bea
\d\la\la\theta|A(t)&=&i\la\la\theta|
[\hat{H}_{\mathrm{S}}(t),A(t)]\d t\nonumber\\
&&-\kappa(t)\la\la\theta|\lp[A(t),\add(t)\}a(t)
+\add(t)[a(t),A(t)\}\rp\d t\nonumber\\
&&-\sigma\big(2\kappa(t)n(t)+\dot{n}(t)\big)\la\la\theta|
[[A(t),\add(t)\},a(t)\}\d t\nonumber\\
&&+\la\la\theta|\lp[A(t),\add(t)\}\d F_t-
\sigma[A(t),a(t)\}\d F^\dag_t\rp,
\label{sde.43}
\eea
where we used
\bea
\la|\d W_t=\la|\d F_t,\quad
\la|\d\tilde{W}_t=\tau^*\la|\d F^\dagger_t.
\eea

The stochastic equation of motion of the Stra\-to\-no\-vich type
for the bra-vector state $\la\la\theta|A(t)$ is derived similarly
in the form
\bea
\d\la\la\theta|A(t)&=&i\la\la\theta|[\hat{H}_{\mathrm{S}}(t),A(t)]\d t
\nonumber\\
&&-\kappa(t)\la\la\theta|\lp\add(t)[a(t),A(t)\}
-\sigma a(t)[\add(t),A(t)\}\rp\d t\nonumber\\
&&+\la\la\theta|\lp[A(t),\add(t)\}\circ\d F_t-
\sigma[A(t),a(t)\}\circ\d F^\dag_t\rp.\label{sde.44}
\eea

Applying to (\ref{sde.43}) the random force ket-vacuum $|\ra$ and the ket-vacuum
$|\mathit{0}\ra$ of the relevant system, one
obtains the equation of motion for the expectation value of an
arbitrary operator $A(t)$ of the relevant system as
\bea
\frac{\d}{\d t}\la\la A(t)\ra\ra&=&
i\la\la[\hat{H}_{\mathrm{S}}(t),A(t)]\ra\ra\nonumber\\
&&-\kappa(t)\la\la\lp[A(t),\add(t)\}a(t)+
\add(t)[a(t),A(t)\}\rp\ra\ra\nonumber\\
&&-\sigma\big(2\kappa(t)n(t)+\dot{n}(t)\big)\la\la
[[A(t),\add(t)\},a(t)\}\ra\ra.
\label{sde.45}
\eea
Here, $\la\la\cdots\ra\ra=\la|\la\theta|\cdots|\mathit{0}\ra|\ra$
means to take both random average and vacuum expectation.
This is the exact equation of motion for systems with
linear-dissipative coupling to reservoir, which can be also
derived by means of the Fokker-Planck equation (\ref{sde.15}).
Here, we used the property
\bea
\la[A(t),\add(t)\}\d F_t\ra=
\la[A(t),a(t)\}\d F_t^\dag\ra=0,
\quad
\eea
which is the characteristics of the It\^o multiplication. Note
that equation of motion for expectation value of an arbitrary
operator $A(t)$ does not depend on the parameter $\lambda$.

\section{Semi-free system with a stationary process}
\label{Stationary}

One possible way to specify a model is to give the Boltzmann
equation (\ref{ndot}). For the cases of a semi-free system
corresponding to the stationary quantum stochastic processes, 
one needs to make substitutions
\bea
i\Sigma^<(t)=2\kappa\bar{n},\quad
\omega(t)=\omega,\quad
\kappa(t)=\kappa,
\eea
where $\bar{n}$ is an average quantum number in equilibrium given by
\bea
\bar{n}&=&\lp\e^{\omega/T}-\sigma\rp^{-1},
\eea
and $T$ is the temperature of environment
(here we use the system with the Boltzmann constant $\kB=1$).
Then, the Boltzmann equation (\ref{ndot}) becomes
\bea
\dot{n}(t)&=&-2\kappa\lp n(t)-\bar{n}\rp.\label{sp.be}
\eea
It describes the system of a damped harmonic oscillator.

Substituting the Boltzmann equation (\ref{sp.be}) into the
semi-free hat-Hamiltonian (\ref{Hs_Pi}) with (\ref{HSt}) and
(\ref{Pit2}) or with (\ref{3.43}) and (\ref{Pit3}), 
one obtains
%
\bea
\hat{H}&=&\omega\lp\ad a-\tad\ta\rp
+2\sigma\tau i\kappa\lp1+\sigma\bar{n}\rp a\ta+
2\sigma\tau i\kappa\bar{n}\ad\tad\nonumber\\
&&{}-i\kappa\lp1+2\sigma\bar{n}\rp\lp\ad a+\tad\ta\rp-2i\kappa\bar{n}
\label{sp.sfhh0}\\
&=&\omega\bar{a}^\mu a^\mu
-i\kappa\bar{a}^\mu A^{\mu\nu}a^\nu
+\sigma(\omega+i\kappa)
\\
&=&\omega\lp\gamma^{\venus}\gamma_t-\tilde{\gamma}^{\venus}\tilde{\gamma}_t\rp
-i\kappa\lp\gamma^{\venus}\gamma_t+\tilde{\gamma}^{\venus}\tilde{\gamma}_t\rp
-2\sigma\tau i\kappa\lp n(t)-\bar{n}\rp\gamma^{\venus}\tilde{\gamma}^{\venus},
\label{sp.sfhh}
\eea
%
where
\bea
A^{\mu\nu}&=&
\lp
\ba{rr}
1+2\sigma\bar{n}&-2\sigma\bar{n}\\
2\lp1+\sigma\bar{n}\rp&-\lp1+2\sigma\bar{n}\rp\\
\ea
\rp.
\eea

The Fokker-Planck equation of the model is given by
\bea
\ddt|\mathit{0}(t)\ra&=&-i\hat{H}|\mathit{0}(t)\ra,\label{sp.evolution}
\eea
with (\ref{sp.sfhh}). It is solved as (\ref{ket1}) with the order parameter
\bea
\label{sp.ket}
\la\theta|\tilde{\gamma}_t\gamma_t|\mathit{0}\ra&=&
\sigma\tau\lp n(0)-\bar{n}\rp\lp1-\e^{-2\kappa t}\rp,
\eea
where $\gamma_t$, $\gamma^{\venus}$ and their tilde conjugates are
defined by (\ref{gamma-bt}) and (\ref{tdbt})
with $n(t)$ being replaced by the solution of (\ref{sp.be}).
The expression (\ref{ket1}) with the order parameter (\ref{sp.ket})
led us to the notion of a mechanism
named the spontaneous creation of dissipation
\cite{Arimitsu87i,Arimitsu87j,Arimitsu88i,Arimitsu87k,Arimitsu87l,Arimitsu88j}.

Introducing a set of new operators
\bea
\bar{d}^\nu=\lp d^\dag,-\tau\tilde{d}\rp,\quad
d^\mu={\mathrm{collon}}\lp d,\tau\tilde{d}^\dag\rp,
\eea
defined by
\bea
\bar{d}^\nu=\bar{a}^\mu \ls B^{\mu\nu}\rs^{-1},\quad
d^\mu=B^{\mu\nu}a^\nu,
\eea
with
\bea
B^{\mu\nu}=\lp
\ba{rr}
1+\sigma\bar{n}&-\sigma\bar{n}\\
-1&1\\
\ea\rp,
\label{B nbar}
\eea
the hat-Hamiltonian $\hat{H}$ can be also written in the form
%
\bea
\hat{H}=\omega\lp d^\dag d-\tilde{d}^\dag\tilde{d}\rp
-i\kappa\lp d^\dag d+\tilde{d}^\dag\tilde{d}\rp\label{sp.sfhhdf}.
\label{sp.sfhhno}
\eea
%
%
We see that the new operators satisfy the canonical $(\mbox{anti-})$commutation relation
\bea
[d^{\mu},\bar{d}^{\nu}]_{-\sigma}=\delta^{\mu\nu},
\eea
and that TSC (\ref{tsc2}) for the thermal ket-vacuum $|\mathit{0}\ra$
%
%
can be expressed as
\bea
\tilde{d}\,|\mathit{0}\ra&=&\tau\lp n(0)-\bar{n}\rp d^\dag|\mathit{0}\ra.
\eea

It is easy to see from the diagonalized form (\ref{sp.sfhhdf}) of
$\hat{H}$ that
%
\bea
d(t)&=&\hat{V}^{-1}(t)~d^{\phantom{\dag}}\hat{V}(t)=
d^{\phantom{\dag}}\e^{-(i\omega+\kappa)t},\\
\tilde{d}^{\dag\!\dag}(t)&=&\hat{V}^{-1}(t)\tilde{d}^\dag\hat{V}(t)
=\tilde{d}^\dag\e^{-(i\omega-\kappa)t}.
\eea
%
On the other hand, it is easy to see from the normal ordered form
(\ref{sp.sfhh}) that $\hat{H}$ satisfies $\la\theta|\hat{H}=0$,
since the annihilation and creation operators satisfy
(\ref{gamma-annihilate Sh}) and (\ref{gamma-annihilate tilde Sh}).
%
%
The difference between the operators which diagonalize $\hat{H}$
and the ones which make $\hat{H}$ in the form of normal product is
one of the features of NETFD, and shows the point that the
formalism is different from usual quantum mechanics and quantum
field theory. This is manifestations of the fact that the
hat-Hamiltonian is a time-evolution generator for irreversible
processes.

The se\-cond law of
thermodynamics tells us that for a closed system the entropy increment $\d\mathcal{S}$
of the relevant system should be given by \cite{Kubo65}
\bea
\d\mathcal{S}&=&\d\mathcal{S}_{\mathrm{i}}+\d\mathcal{S}_{\mathrm{e}},
\label{sp.entropychange}\\
\d\mathcal{S}_{\mathrm{i}}&\geq &0,
\label{sp.entropychange1}
\eea
where $\d\mathcal{S}_{\mathrm{i}}$ is the change of intrinsic
entropy of the system, and $\d\mathcal{S}_{\mathrm{e}}$ the change
due to the heat flow $\dbar Q$ into the system from the thermal reservoir with 
temperature $T$:
\bea
\d\mathcal{S}_{\mathrm{e}}&=&\frac{\dbar Q}{T}.
\eea

We can check this for the present model \cite{Arimitsu94}.
The entropy of the relevant system is given by \cite{Balescu75}
\bea
\mathcal{S}(t)=-\lc n(t)\ln n(t)
-\sigma\ls 1+\sigma n(t)\rs \ln\ls 1+\sigma n(t)\rs\rc,
\quad\;\label{sp.entropy}
\eea
whereas the heat change of the system can be identified with
\bea
\dbar Q(t)&=&\omega\,\d n(t)
\eea
leading to
\bea
\d\mathcal{S}_{\mathrm{e}}&=&\frac{\omega}{T}\d n(t).
\label{sp.heatchange}
\eea
Putting (\ref{sp.entropy}) and (\ref{sp.heatchange}) into
(\ref{sp.entropychange}) for $\d\mathcal{S}$ and
$\d\mathcal{S}_{\mathrm{e}}$, respectively, we have a relation for
the entropy production rate \cite{Arimitsu94}
\bea
\frac{\d\mathcal{S}_{\mathrm{i}}}{\d t}=
\frac{\d\mathcal{S}}{\d t}-\frac{\d\mathcal{S}_{\mathrm{e}}}{\d t}
=2\kappa\big(n(t)-\bar{n}\big)
\ln\frac{n(t)(1+\sigma\bar{n})}{\bar{n}\big(1+\sigma n(t)\big)}
\geq 0,
\eea
giving the inequality (\ref{sp.entropychange1}).
It is easy to check that inequality holds for both cases, i.e.
$n(t)>\bar{n}$ and $n(t)<\bar{n}$, and that the equality realizes
for the thermal equilibrium state, $n(t)=\bar{n}$, or for the
quasi-static process with $\kappa\to0$.

\section{Relation to the Monte Carlo wave-function method}
\label{MC-WF}

In this section, we will investigate the Fokker-Planck equation
(\ref{sp.evolution}) in order to reveal the relation of NETFD to
the Monte Carlo wave-function method, i.e. the quantum jump simulation
\cite{Dalibard92,Molmer93,Molmer96,Gardiner92,Dum92,Carmichael93,Garraway94,Plenio98,Breuer02}
in which evolution with a non-hermitian hat-Hamiltonian is
described in terms of randomly decided quantum jumps followed by
the wave-function normalization.

Let us decompose the hat-Hamiltonian (\ref{sp.sfhh0}) as
\bea
\hat{H}&=&\hat{H}^{(0)}+\hat{H}^{(1)},
\eea
with
%
\bea
\hat{H}^{(0)}&=&\omega\lp a^\dag a-\tilde{a}^\dag\tilde{a}\rp
-i\kappa(1+2\sigma\bar{n})\lp a^\dag a+\tilde{a}^\dag\tilde{a}\rp,\qquad\\
\hat{H}^{(1)}&=&2i\sigma\tau\kappa\lp(1+\sigma\bar{n})a\tilde{a}
+\bar{n}a^\dag\tilde{a}^\dag\rp-2i\kappa\bar{n},
\eea
%
and consider an equation:
\bea
\ddt|\mathit{0}_0(t)\ra'&=&-i\hat{H}^{(0)}|\mathit{0}_0(t)\ra'.
\label{F-P prime}
\eea
Note that $\hat{H}^{(1)}$ contains cross terms among tilde and
non-tilde operators. We see that $\hat{H}^{(0)}$ and $\hat{H}^{(1)}$ have
the properties
%
\bea
\la\theta|\hat{H}^{(0)}&=&-2i\kappa(1+2\sigma\bar{n})\la\theta|a^\dag a ,\\
\la\theta|\hat{H}^{(1)}&=&2i\kappa(1+2\sigma\bar{n})\la\theta|a^\dag a.
\eea

Introducing the wave-functions $|\psi(t)\ra$ and
$|\tilde{\psi}(t)\ra$ thro\-ugh the relation
\bea
|\mathit{0}_0(t)\ra'&=&|\psi(t)\ra|\tilde{\psi}(t)\ra,
\eea
we have from (\ref{F-P prime}) the Schr\"odinger equations of the
form
\bea
\ddt|\psi(t)\ra&=&-iH^{(0)}|\psi(t)\ra,
\label{Sch eq prime}
\eea
and its tilde conjugate, where
\bea
H^{(0)}&=&\omega a^\dag a-i\kappa(1+2\sigma\bar{n})a^\dag a.
\eea
This procedure is possible because $\hat{H}^{(0)}$ does not contain
cross terms among tilde and non-tilde operators. The Monte Carlo
simulations for quantum systems are performed for the
Schr\"odinger equation (\ref{Sch eq prime})
\cite{Dalibard92,Molmer93,Molmer96,Gardiner92}.

The time evolution generated by the hat-Hamiltonian $\hat{H}^{(0)}$ does
not preserve the normalization of the ket-vacuum, i.e. the
normalized ket-vacuum $|\mathit{0}(t)\ra$ evolves for the time
increment $\d t$ as
\bea
\la\theta|\mathit{0}_0(t+\d t)\ra'=\la\theta|(1-i\hat{H}^{(0)}\d t)
|\mathit{0}(t)\ra
=1-\d p(t),
\eea
with
\bea
\d p(t)= i \la\theta| \hat{H}^{(0)} |\mathit{0}(t)\ra dt
=2\kappa(1+2\sigma\bar{n})n(t)\d t.
\eea

The recipe of the quantum jump simulation is that, for a time
increment $\d t$,
\begin{enumerate}
\item[1)]
when $\d p(t)<\varepsilon$ with a given positive constant
$\varepsilon$, the normalized ket-vacuum evolves as
\bea
|\mathit{0}(t)\ra\rightarrow|\mathit{0}_0(t+\d t)\ra&=&
\frac{|\mathit{0}_0(t+\d t)\ra'}{1-\d p(t)}\nonumber\\
&=&\frac{|\psi(t+\d t)\ra}{\sqrt{1-\d p(t)}}
\frac{|\tilde{\psi}(t+\d t)\ra}{\sqrt{1-\d p(t)}};\qquad\;
\eea
\item[2)]
in the case $\d p(t)>\varepsilon$, a quantum jump comes in
\bea
|\mathit{0}_1(t+\d t)\ra=\frac{-i\hat{H}^{(1)}\d t|0(t)\ra}{\d p(t)}.
\eea
\end{enumerate}
The time increment $\d t$ should be chosen as the condition $\d
p(t)\ll1$ being satisfied.

Averaging the processes $|\mathit{0}_0(t)\ra$ and $|\mathit{0}_1(t)\ra$ 
with respective probabilities $1-\d p(t)$ and $\d p(t)$:
\bea
|\mathit{0}(t+\d t)\ra=[1-\d p(t)]|\mathit{0}_0(t+\d t)\ra
+ \d p(t)|\mathit{0}_1(t+\d t)\ra,
\eea
we can obtain the Fokker-Planck equation (\ref{sp.evolution}).
Note that the ket-vacuums $|\mathit{0}_0(t)\ra$ and $|\mathit{0}_1(t)\ra$
look like satisfying a certain kind of stochastic Liouville
equation.

\section{Summary}
\label{Conclusions}

The aim of this paper has been to study the system of QSDEs 
from a physical basis. We have
formulated everything from the starting point using the method of
NETFD. In the presented approach, boson and fermion systems are
considered simultaneously. The obtained results have
two fixed parameters: the real parameter $\sigma$
specifying different commutation rules for boson and fermion
operators, and the complex parameter $\tau$, (\ref{tau}), specifying
different thermal state conditions for boson and fermion systems.
Such a combined consideration was made possible due to the
unification of fermion and boson stochastic calculus (Appendix
\ref{bfbm}), where fermion annihilation and creation processes are
realized in a Boson Fock space by means of a simple stochastic
integral prescription leading to similar multiplication rules for
stochastic differentials.

The dissipation mechanism is considered through the
concept of a quantum noise, i.e. as a quantum field interacting
with the relevant system. In our paper we considered two types of
interaction with external fields: hermitian ($\lambda=1$) and 
non-hermitian ($\lambda = 0$).
With the latter, conservation of the probability is satisfied
within the relevant system. With the former, information about
only relevant system is not enough and instead of that we can
speak about conservation of the probability within the total
system: relevant system plus environment system.

As we are concentrated on the stochastic equations, there are two
types of stochastic calculus: It\^o and Stratonovich.
Correspondingly, equations used one or another type of stochastic
calculus are classified as QSDE of the It\^o or Stratonovich
types. The Langevin equation of the Stratonovich type
(\ref{sde.29}) has structure similar to one of the Heisenberg
equation of motion for a dynamical quantity in quantum mechanics
and quantum field theory. As a result of different stochastic
multiplication rule, the Langevin equation of the It\^o type
(\ref{sde.38}) contains an extra term proportional to a product of
random forces $\d W_t\d\tilde{W}_t$. The corresponding
Fokker-Planck equation is then obtained most easily from the quantum
stochastic Liouville equation of the It\^o type by taking the
random average. Though in fermion case the connection with the
classical Brownian motion is only formal, the It\^o/Stratonovich
product formula is the same as in boson case (relations
(\ref{IS.6}) and (\ref{IS.7})). The averaged equation of motion
for a dynamical quantity can be obtained in two ways. From the
Langevin equation by taking both random average and the relevant
vacuum expectation, or from the Fokker-Planck equation by taking
the vacuum expectation of operators corresponding to the dynamical
quantity. In our study we showed that QSDEs constructed upon
hermitian and non-hermitian interaction hat-Hamiltonians lead to
the same averaged equation of motion (\ref{sde.45}) for an
arbitrary operator of the relevant system. In the case of
stationary semi-free quantum stochastic process, its
irreversibility is checked in terms of the Boltzmann entropy. We
also demonstrated the relationship between the presented
formulation and the method of quantum jump simulations.

The approach we followed in this article is rather formal and
we are looking now for some demonstrative examples of its
application for particular problems. An interesting result is
obtained, for instance, for the model of a continuous quantum
non-demolition measurement --- continuous observation of a
particle track in the cloud chamber
\cite{Arimitsu99a,Arimitsu02b}, and for the system
cor\-res\-pon\-ding to the quantum Kra\-mers equation
\cite{Arimitsu99b,Arimitsu02a}. More detailed report about them
will be presented elsewhere.

\section*{Acknowledgments}

\noindent
Authors would like to thank Mr.~Y.~Fukuda and Mr.~Y.~Kaburaki for their
fruitful discussions.

\appendix

\section{It\^o and Stratonovich calculus}
\label{IS}

Definitions of the It\^o \cite{Ito44} and Stratonovich
\cite{Stratonovich66} mul\-ti\-pli\-ca\-ti\-ons for arbitrary
stochastic operators $X_t$ and $Y_t$ in the Schr\"odinger
representation are given, respectively, by
%
\bea
X_t\cdot\d Y_t&=&X_t(Y_{t+\d t}-Y_t),\label{IS.1sa}\\
\d X_t\cdot Y_t&=&(X_{t+\d t}-X_t)Y_t,\label{IS.1sb}
\eea
%
and
%
\bea
X_t\circ\d Y_t&=&\frac12(X_{t+\d t}+X_t)(Y_{t+\d t}-Y_t),\label{IS.2sa}\\
\d X_t\circ Y_t&=&(X_{t+\d t}-X_t)\frac12(Y_{t+\d t}+Y_t).\label{IS.2sb}
\eea
%
From these relations we have the connection formulae between the
It\^o and Stratonovich products in the differential form as
%
\bea
\ds X_t\circ\d Y_t&=&\ds X_t\cdot\d Y_t+\frac12\d X_t\cdot\d Y_t,
\label{IS.4a}\\
\ds\d X_t\circ Y_t&=&\ds\d X_t\cdot Y_t+\frac12\d X_t\cdot\d Y_t.
\label{IS.4b}
\eea
%
Note that random average of the stochastic multiplication 
(\ref{IS.1sa}) or (\ref{IS.1sb})
of the It\^o type is equal to zero.

Definitions of the It\^o and Stratonovich
mul\-ti\-pli\-ca\-ti\-ons for stochastic operators $X(t)$ and
$Y(t)$ in the Heisenberg representation are given in the same form
by
%
\bea
X(t)\cdot\d Y(t)&=&X(t)\ls Y(t+\d t)-Y(t)\rs,\label{IS.1a}\\
\d X(t)\cdot Y(t)&=&\ls X(t+\d t)-X(t)\rs Y(t),\label{IS.1b}
\eea
%
and
%
\bea
X(t)\circ\d Y(t)&=&\frac12 \ls X(t+\d t)+X(t)\rs
\ls Y(t+\d t)-Y(t)\rs,\label{IS.2a}\\
\d X(t)\circ Y(t)&=&\ls X(t+\d t)-X(t)\rs\frac12\ls
Y(t+\d t)+Y(t)\rs,\label{IS.2b}
\eea
%
where operators $X(t)$ and $\d X(t)$ are introduced, respectively,
through relations
%
\bea
X(t)&=&\hat{V}_F^{-1}(t)X_t\hat{V}_F(t),\label{IS.5a}\\
\d X(t)&=&\d(\hat{V}_F^{-1}(t)X_t\hat{V}_F(t)),\label{IS.5b}
\eea
%
with $\hat{V}_F(t)$ being a stochastic time evolution operator.

From (\ref{IS.1a}) to (\ref{IS.2b}), 
we have the connection formulae
between the It\^o and Stratonovich products in the differential
form as
%
\bea
X(t)\circ\d Y(t)&=&\ds X(t)\cdot\d Y(t)
+\frac12\d X(t)\cdot\d Y(t),\label{IS.3a}\\
\d X(t)\circ Y(t)&=&\ds\d X(t)\cdot Y(t)
+\frac12\d X(t)\cdot\d Y(t).\qquad\label{IS.3b}
\eea
%

Stochastic multiplications (\ref{IS.1a}) to (\ref{IS.2b}) are
consistent with corresponding types of differential calculus for
products of stochastic operators, which for the case of the It\^o
type calculus and the Stratonovich type calculus read,
respectively, as
\bea
\d[X(t)Y(t)]=\d X(t)\cdot Y(t)+X(t)\cdot\d Y(t)+\d X(t)\cdot\d Y(t),
\label{IS.6}
\eea
and
\bea
\d[X(t)Y(t)]&=&\d X(t)\circ Y(t)+X(t)\circ\d Y(t).
\qquad\qquad\label{IS.7}
\eea

\section{Boson and fermion Brownian motion}
\label{bfbm}

Let $\Gamma^0_{\mathrm{s}}$ denotes the boson Fock space (the
symmetric Fock space) over the Hilbert space
$\mathscr{H}=L^2(\mathbf{R}_+)$ of square integrable functions,
and $b_t$ and $b_t^\dag$ denote, respectively, boson annihilation
and creation operators at time $t\in[0,\infty)$ satisfying the
canonical commutation relations
\begin{eqnarray}
\label{boson-commutators}
[b_t,b^\dag_s]=\delta(t-s),\quad[b_t,b_s]=[b^\dag_t,b^\dag_s]=0.
\end{eqnarray}
The bra- and ket-vacuums $(|$ and $|)$ are defined, respectively, by
\begin{eqnarray}
(|b^\dag_t=0,&\quad&
b_t|)=0.\label{b2}
\end{eqnarray}
Note that $(|=|)^\dag$ since here we are considering the unitary
representation of $b_t$ and $b_t^\dag$. The space
$\Gamma^0_{\mathrm{s}}$ is equipped with a total family of
exponential vectors
%
\begin{eqnarray}
(e(f)|&=&(|\exp\lc\int_0^\infty\d t\;f^*(t)b_t\rc,\\
|e(\g))&=&\exp\lc\int_0^\infty\d t\;\g(t)b^\dag_t\rc|),
\label{exp-vectors}
\end{eqnarray}
%
whose overlapping is
\begin{eqnarray}
(e(f)|e(\g))&=&\exp\lc\int_0^\infty\d t\;f^*(t)\g(t)\rc.
\end{eqnarray}
Here, $f$, $g\in\mathscr{H}$. The dense span of exponential vectors is
denoted by $\mathscr{E}$. Operators $b_t$, $b^\dag_t$ and
exponential vectors are characterized by the relations
\begin{eqnarray}
\label{d5}
(e(f)|b^\dag_t=(e(f)|f^*(t),\quad
b_t|e(\g))=\g(t)|e(\g)).
\end{eqnarray}

Let us introduce operator $U_t$ defined as
\begin{eqnarray}
U_t=\sigma_<P_{[0,t]}+\sigma_>P_{(t,\infty)},
\end{eqnarray}
where $\sigma_<$ and $\sigma_>$ are two independent parameters
taking values $\pm1$, and $P_{[a,b]}$ ($a$$\leq $$b$) is an
operator on $\mathscr{H}$ of multiplication by the indicator
function whose action reads
\begin{eqnarray}
P_{[a,b]}\int_0^\infty\d t\;\g(t)=\int_a^b\d t\;\g(t)
=\int_0^\infty\d t\;\theta(t-a)\theta(b-t)\g(t).\label{Pab}\qquad
\end{eqnarray}
Here, $\theta(t)$ is the step function specified by
\begin{eqnarray}
\theta(t)=\lc
\ba{cl}
1&\mbox{for}\;t\geq 0,\\
0&\mbox{for}\;t<0.
\ea
\right.
\end{eqnarray}
The operator $P_{[a,b]}$ has the following properties:
\begin{equation}
P_{[a,b]}^2=P_{[a,b]},\quad
P_{[a,b]}^\dag=P_{[a,b]},\quad
P_{[a,b]}P_{[c,d]}=P_{[c,d]}P_{[a,b]},
\end{equation}
which are easily verified using the definition (\ref{Pab}). Then, we
see that operator $U_t$ is unitary, and satisfies
\begin{eqnarray}
\label{properties-of-U}
U_t^2=I,\quad
U_t^\dag=U_t,\quad
U_tU_s=U_sU_t,
\end{eqnarray}
where $I$ is the identity operator. 

The so-called
\textit{reflection process} $J_t\equiv J_t(U_t)$,
$t\in\mathbf{R}_+$, whose action on $\mathscr{E}$ is given by
\cite{Hudson86}
\bea
J_t|e(\g))=|e(U_t\g))
=\exp\lc U_t\int_0^\infty\d t'\;\g(t')b_{t'}^\dag\rc|),
\quad\label{Jte}
\eea
inherits properties of the operator $U_t$, (\ref{properties-of-U}), i.e.
\begin{eqnarray}
\label{b12b}
J_t^2=\textbf{1},\quad
J_t^\dag=J_t,\quad
J_tJ_s=J_sJ_t,
\end{eqnarray}
and does not change the vacuum:
\begin{eqnarray}
(|J_t=(|,&\quad&J_t|)=|).
\end{eqnarray}
Here, $\textbf{1}$ is the unit operator defined in $\Gamma^0_{\mathrm{s}}$.

Let us now consider new operators
\begin{eqnarray}
\mathsf{b}_t=J_tb_t,&\quad&
\mathsf{b}_t^\dag=b_t^\dag J_t.\label{JbbJ}
\end{eqnarray}
Apparently, they annihilate vacuums
\begin{eqnarray}
(|\mathsf{b}_t^\dag=0,&\quad& \mathsf{b}_t|)=0.
\end{eqnarray}
The following matrix elements
%
\begin{eqnarray}
(e(f)|[J_t,b_s]_{-\sigma}|e(\g))&=&
(e(f)|\{1\nonumber\\
&&-\sigma[\sigma_>
+(\sigma_<-\sigma_>)
\theta(t-s)]\}\g(s)J_t|e(\g)),\\
(e(f)|[\mathsf{b}_t,\mathsf{b}_s^\dag]_{-\sigma}|e(\g))&=&
(e(f)|\delta(t-s)|e(\g))\nonumber\\
&&+(e(f)|J_tJ_s(\sigma_>\sigma_<-\sigma)f^*(s)\g(t)|e(\g)),
\end{eqnarray}
%
are valid for $f$, $g\in\mathscr{H}$. Then the requirement of equal-time
(anti-)\-com\-mu\-ta\-ti\-vi\-ty between $J_t$ and $b_t$
\bea
[J_t,b_t]_{-\sigma}=0
\eea
gives
\begin{eqnarray}
1-\sigma\sigma_<=0,\label{Jtbt}
\end{eqnarray}
while the requirement of canonical (anti-)commutation relation
\begin{eqnarray}
[\mathsf{b}_t,\mathsf{b}_s^\dag]_{-\sigma}&=&\delta(t-s)\label{ctcs}
\end{eqnarray}
leads to
\begin{eqnarray}
\sigma_>\sigma_<-\sigma=0.\label{btbs}
\end{eqnarray}
All those conditions are satisfied when $\sigma_<=\sigma$ and $\sigma_>=+1$.
Then the operator $U_t$ turns out to be
\begin{eqnarray}
U_t&=&\sigma P_{[0,t]}+P_{(t,\infty)}.
\end{eqnarray}
Note that for a boson system, i.e. $\sigma=1$, $U_t=I$ and the
operators $\mathsf{b}_t$ and $\mathsf{b}_t^\dag$ reduce,
respectively, to $b_t$ and $b_t^\dag$.

We see that the generalized quantum Brownian motion, defined by
\begin{eqnarray}
\mathsf{B}_t=\int_0^t\d t'\;\mathsf{b}_{t'},
&\qquad&
\mathsf{B}_t^\dag=\int_0^t\d t'\;\mathsf{b}^\dag_{t'},
\label{b13}
\end{eqnarray}
with $\mathsf{B}_0=0$, $\mathsf{B}^\dag_0=0$, satisfies
\begin{eqnarray}
[\mathsf{B}_t,\mathsf{B}_s^\dag]_{-\sigma}&=&\min(t,s).
\end{eqnarray}
The case $\sigma=1$ represents the \textit{boson Brownian motion}
\cite{Hudson84a,Parthasarathy85}, whereas the case $\sigma=-1$ the
\textit{fermion Brownian motion} \cite{Hudson86}. Their increments
%
\begin{eqnarray}
\d\mathsf{B}_t&=&\mathsf{B}_{t+\d t}-\mathsf{B}_t^{\phantom{\dag}}
=\mathsf{b}_t\d t,\\
\d\mathsf{B}_t^\dag&=&\mathsf{B}_{t+\d t}^\dag-\mathsf{B}_t^\dag
=\mathsf{b}_t^\dag\d t,
\end{eqnarray}
%
annihilate the vacuum, i.e.
\begin{eqnarray}
(|\d\mathsf{B}^\dag_t=0,&\quad&\d\mathsf{B}_t|)=0,
\end{eqnarray}
and their matrix elements read
%
\begin{eqnarray}
(e(f)|\d\mathsf{B}_t|e(\g))&=&(e(f)|J_t\g(t)\d t|e(\g)),\\
(e(f)|\d\mathsf{B}_t^\dag|e(\g))&=&(e(f)|f^*(t)\d tJ_t|e(\g)),\qquad\\
(e(f)|\d\mathsf{B}_t\d\mathsf{B}_t|e(\g))&=&0,\\
(e(f)|\d\mathsf{B}_t^\dag\d\mathsf{B}_t|e(\g))&=&0,\\
(e(f)|\d\mathsf{B}_t\d\mathsf{B}_t^\dag|e(\g))&=&\d t(e(f)|e(\g)).
\end{eqnarray}
%
Here we neglected terms of the higher order than $\d t$. The
latter equations are summarized in the following table of
multiplication rules for increments $\d\mathsf{B}_t$ and
$\d\mathsf{B}_t^\dag$:
\bea
\label{dBdBrule}
\ba{l|lll}
&\d\mathsf{B}_t&\d\mathsf{B}_t^\dag&\d t\\\hline
\d\mathsf{B}_t&0&\d t&0\\
\d\mathsf{B}_t^\dag&0&0&0\\
\d t&0&0&0\\
\ea
\eea

Now we consider a tensor product space
$\hat\Gamma=\Gamma_s^0\otimes\tilde{\Gamma}_s^0$. Its vacuum
states $|)\!)$ and exponential vectors $|e(f,\g))\!)$ are defined
through the ``principle of correspondence'' \cite{Arimitsu87a}
%
\bea
|)\!)&\longleftrightarrow&|)(|,\\
|e(f,\g))\!)&\longleftrightarrow&|e(f))(e(\g)|.
\eea
%
Annihilation and creation operators acting on $\hat{\Gamma}$ are
defined through
%
\bea
b_t|e(f,\g))\!)&\longleftrightarrow&b_t|e(f))(e(\g)|,\\
b_t^\dag|e(f,\g))\!)&\longleftrightarrow&b_t^\dag|e(f))(e(\g)|,\\
\tilde{b}_t|e(f,\g))\!)&\longleftrightarrow&|e(f))(e(\g)|b_t^\dag,\\
\tilde{b}_t^\dag|e(f,\g))\!)&\longleftrightarrow&|e(f))(e(\g)|b_t,
\eea
%
and similarly for $J_t$ and $\tilde{J}_t$, i.e.
%
\bea
J_t|e(f,\g))\!)&\longleftrightarrow&J_t|e(f))(e(\g)|,\\
\tilde{J}_t|e(f,\g))\!)&\longleftrightarrow&|e(f))(e(\g)|J_t.
\eea
%
Algebra of commutation relations between these operators reads
%
\bea
[b_t,b_s^\dag]&=&[\tilde{b}_t,\tilde{b}_s^\dag]=\delta(t-s),\\\relax
[b_t,\tilde{b}_s]&=&[b_t,\tilde{b}_s^\dag]=0,\\\relax
[J_t,\tilde{b}_s]&=&[\tilde{J}_t,b_s]=0,\\\relax
[J_t,b_t]_{-\sigma}&=&[\tilde{J}_t,\tilde{b}_t]_{-\sigma}=0.
\eea
%

Let us now consider new operators defined by
%
\bea
\mathsf{b}_t &=& J_tb_t,\quad \mathsf{b}_t^\dag = b_t^\dag J_t,\label{no-tilde-fb}\\
\tilde{\mathsf{b}}_t &=& \hat\tau\tilde{J}_t\tilde{b}_t,\quad
\tilde{\mathsf{b}}_t^\dag = \hat\tau\tilde{b}_t^\dag\tilde{J}_t,
\label{tilde-fbt}
\eea
%
where $\hat\tau$ is an operator satisfying the following
(anti-)\-com\-mu\-ta\-ti\-on relations
%
\bea
[\hat\tau,J_t]&=&[\hat\tau,\tilde{J}_t]=0,\\\relax
[\hat\tau,b_t]_{-\sigma}&=&[\hat\tau,b_t^\dag]_{-\sigma}=0,\\\relax
[\hat\tau,\tilde{b}_t]_{-\sigma}&=&[\hat\tau,\tilde{b}_t^\dag]_{-\sigma}=0,
\eea
%
and the condition
\bea
\hat\tau^2=\sigma.\label{hat-tau-2}
\eea
Operators $\mathsf{b}_t$, $\mathsf{b}_t^\dag$ and their tilde
conjugates annihilate vacuums
\bea
(\!(|\mathsf{b}_t^\dag=(\!(|\tilde{\mathsf{b}}_t^\dag=0,\quad
\mathsf{b}_t|)\!)=\tilde{\mathsf{b}}_t|)\!)=0,
\eea
and satisfy canonical (anti-)commutation relations
%
\bea
[\mathsf{b}_t,\mathsf{b}_s^\dag]_{-\sigma}&=&
[\tilde{\mathsf{b}}_t,\tilde{\mathsf{b}}_s^\dag]_{-\sigma}=\delta(t-s),
\label{req-fbt}\\\relax
[\mathsf{b}_t,\tilde{\mathsf{b}}_s]_{-\sigma}&=&
[\mathsf{b}_t,\tilde{\mathsf{b}}_s^\dag]_{-\sigma}=0.
\label{tau-condition-1}
\eea
%
Since $(\tilde{\mathsf{b}}_t)^\dag$ and
$(\mathsf{b}_t^\dag)^\sim$ are calculated as
\bea
(\tilde{\mathsf{b}}_t)^\dag&=&\tilde{b}_t^\dag\tilde{J}_t\hat\tau^\dag
=\sigma\hat\tau^\dag\tilde{b}_t^\dag\tilde{J}_t,\\
(\mathsf{b}_t^\dag)^\sim&=&\hat\tau(b_t^\dag J_t)^\sim
=\hat\tau\tilde{b}_t^\dag\tilde{J}_t,
\eea
the commutativity of tilde conjugation and hermitian conjugation for
operators (\ref{no-tilde-fb}) and (\ref{tilde-fbt}) implies
\bea
\hat\tau^\dag=\sigma\hat\tau.\label{hat-tau-dag}
\eea
In order to fulfill the requirement that double tilde conjugation
applied to operators $\mathsf{b}_t$'s leaves them unchanged one
needs to put
\bea
(\hat\tau)^\sim=\hat\tau^\dag,
\eea
since
\bea
(\tilde{\mathsf{b}}_t)^\sim&=&\hat\tau(\hat\tau)^\sim J_tb_t
=\hat\tau(\hat\tau)^\sim\mathsf{b}_t.
\eea
Because of
\bea
\tilde{\mathsf{b}}_t|)\!)
=\tilde{b}_t\Big(\hat{\tau}|)\!)\Big)=0,
\eea
one can conclude that $\hat{\tau}|)\!)\propto|)\!)$. The
proportionality factor is a phase factor since the norm of
$\hat{\tau}|)\!)$ is unity:
\bea
(\!(|\hat{\tau}^\dag\hat{\tau}|)\!)=(\!(|\sigma\hat{\tau}\hat{\tau}|)\!)
=\sigma^2=1.
\eea
Hence one can write
\bea
\hat{\tau}|)\!)=\e^{i\phi/2}|)\!).
\eea
Multiplying both sides by $\hat{\tau}$, one has
\bea
\hat{\tau}^2|)\!)=\e^{i\phi}|)\!)=\sigma|)\!),
\eea
which gives $\e^{i\phi}=\sigma$, or
\bea
\hat{\tau}|)\!)&=&\sqrt{\sigma}\;|)\!).
\eea

Thermal degree of freedom can be introduced by the Bogoliubov
transformation in $\hat\Gamma$. For this purpose we re\-qui\-re
that the expectation value of $\mathsf{b}_t^\dag\mathsf{b}_s$
should be
\begin{eqnarray}
\la\mathsf{b}_t^\dag\mathsf{b}_s\ra=\bar{n}\delta(t-s)\label{req}
\end{eqnarray}
with $\bar{n}\in\mathbf{R}_+$, where $\la\cdots\ra=\la|\cdots|\ra$
indicates the expectation with respect to tilde invariant thermal
vacuums $\la|$ and $|\ra$. The requirement (\ref{req}) is
consistent with TSC for states $\la|$ and $|\ra$ such that
\begin{eqnarray}
\la|\tilde{\mathsf{b}}_t^\dag=\tau^*\la|\mathsf{b}_t,&\quad&
\tilde{\mathsf{b}}_t|\ra=\frac{\tau\bar{n}}{1+\sigma\bar{n}}\mathsf{b}_t^\dag|\ra.
\label{TSC-b}
\end{eqnarray}

Let us introduce annihilation and creation operators
%
\begin{eqnarray}
\mathsf{c}_t&=&[1+\sigma\bar{n}]\mathsf{b}_t-\sigma\tau\bar{n}\tilde{\mathsf{b}}_t^\dag,
\label{aco-c0}\\
\tilde{\mathsf{c}}_t^{\venus}&=&\tilde{\mathsf{b}}_t^\dag-\sigma\tau\mathsf{b}_t,
\label{aco-c}
\end{eqnarray}
%
and their tilde conjugates. From the TSC (\ref{TSC-b}) one has
\begin{eqnarray}
\la|\mathsf{c}_t^{\venus}=\la|\tilde{\mathsf{c}}_t^{\venus}=0,&\quad&
\mathsf{c}_t|\ra=\tilde{\mathsf{c}}_t|\ra=0.
\end{eqnarray}
With the thermal doublet notations
\begin{eqnarray}
\bar{\mathsf{b}}_t^\mu=\lp\mathsf{b}_t^\dag,-\tau\tilde{\mathsf{b}}_t\rp,\quad
\mathsf{b}_t^\nu=\mathrm{collon}\lp\mathsf{b}_t,
\tau\tilde{\mathsf{b}}_t^\dag\rp,
\end{eqnarray}
and
\begin{eqnarray}
\bar{\mathsf{c}}_t^\mu=\lp\mathsf{c}_t^{\venus},-\tau\tilde{\mathsf{c}}_t\rp,\quad
\mathsf{c}_t^\nu=\mathrm{collon}\lp\mathsf{c}_t,
\tau\tilde{\mathsf{c}}_t^{\venus}\rp,
\end{eqnarray}
(\ref{aco-c0}), (\ref{aco-c}) and their tilde conjugates can be written in form of
the Bogoliubov transformation
\begin{eqnarray}
\mathsf{c}_t^\mu=B^{\mu\nu}\mathsf{b}_t^\nu,\quad
\bar{\mathsf{c}}_t^\nu=\bar{\mathsf{b}}_t^\mu[B^{-1}]^{\mu\nu},
\end{eqnarray}
with (\ref{B nbar}).
%
%
This new operators satisfy 
the canonical (anti-)\-com\-mu\-ta\-ti\-on relations
\bea
[\mathsf{c}_t,\mathsf{c}_s^{\venus}]_{-\sigma}=\delta(t-s).
\eea

In the following, we will use the representation space constructed
on vacuums $\la|$ and $|\ra$. Note that $\la|\ne|\ra^\dag$, i.e.
it is not a unitary representation. Let $\hat\Gamma^\beta$ denotes
the Fock space spanned by the basic bra- and ket-vectors
introduced by a cyclic operations of $\mathsf{c}_t$,
$\tilde{\mathsf{c}}_t$ on the thermal bra-vacuum $\la|$, and of
$\mathsf{c}_t^{\venus}$, $\tilde{\mathsf{c}}_t^{\venus}$ on
the thermal ket-va\-cu\-um~$|\ra$. Quantum Brownian motion at
finite temperature is defined in the Fock space $\hat\Gamma^\beta$
by operators
\begin{eqnarray}
\mathsf{B}_t^\sharp=\int_0^t\d s\;\mathsf{b}_s^\sharp,&\quad&
\tilde{\mathsf{B}}_t^\sharp=\int_0^t\d s\;\tilde{\mathsf{b}}_s^\sharp,
\end{eqnarray}
with $\mathsf{B}_0^\sharp=0$ and
$\tilde{\mathsf{B}}_0^\sharp=0$, where $\sharp$ stands for null
or dagger. The explicit representation of processes
$\mathsf{B}_t^\sharp$ and $\tilde{\mathsf{B}}_t^\sharp$ can be
performed in terms of the Bogoliubov transformation. The couple
$\mathsf{B}_t$ and $\mathsf{B}_t^\dag$, for example, is
calculated~as
\begin{eqnarray}
\mathsf{B}_t&=&\int_0^t\d s\;(\mathsf{c}_s
+\sigma\tau\bar{n}\tilde{\mathsf{c}}_s^{\venus})
=\mathsf{C}_t+\sigma\tau\bar{n}\tilde{\mathsf{C}}_t^{\venus},\\
\mathsf{B}_t^\dag&=&\int_0^t\d s\;([1+\sigma\bar{n}]\mathsf{c}_s^{\venus}
+\tau\tilde{\mathsf{c}}_s)
=[1+\sigma\bar{n}]\mathsf{C}_t^{\venus}+\tau\tilde{\mathsf{C}}_t,
\end{eqnarray}
where we defined new operators
\begin{eqnarray}
\mathsf{C}_t^\sharp=\int_0^t\d s\;\mathsf{c}_s^\sharp,&\quad&
\tilde{\mathsf{C}}_t^\sharp=\int_0^t\d s\;\tilde{\mathsf{c}}_s^\sharp,
\end{eqnarray}
with $\mathsf{C}_0^\sharp=0$ and
$\tilde{\mathsf{C}}_0^\sharp=0$, and $\sharp$ standing for
null or the Venus-mark.
Since matrix elements of $\d\mathsf{C}_t^\sharp$ and
$\d\tilde{\mathsf{C}}_t^\sharp$ in thermal space
$\hat\Gamma^\beta$ read
%
\begin{eqnarray}
\la\d\mathsf{C}_t\ra=\la\d\tilde{\mathsf{C}}_t\ra&=&
\la\d\mathsf{C}_t^{\venus}\ra=\la\d\tilde{\mathsf{C}}_t^{\venus}\ra=0,\\
\la\d\mathsf{C}_t^{\venus}\d\mathsf{C}_t\ra&=&
\la\d\tilde{\mathsf{C}}_t^{\venus}\d\tilde{\mathsf{C}}_t\ra=0,\\
\la\d\mathsf{C}_t\d\mathsf{C}_t^{\venus}\ra&=&
\la\d\tilde{\mathsf{C}}_t\d\tilde{\mathsf{C}}_t^{\venus}\ra=\d t,
\end{eqnarray}
%
calculation of moments of quantum Brownian motion in the thermal
space $\hat\Gamma^\beta$ can be performed, for instance, as
\begin{eqnarray}
\la\d\mathsf{B}_t\d\mathsf{B}_t^\dag\ra&=&\La\!\lp\d\mathsf{C}_t
+\sigma\tau\bar{n}\d\tilde{\mathsf{C}}_t^{\venus}\rp
\!\!\lp[1+\sigma\bar{n}]\d\mathsf{C}_t^{\venus}
+\tau\d\tilde{\mathsf{C}}_t\rp\!\Ra\nonumber\\
&=&[1+\sigma\bar{n}]\la\d\mathsf{C}_t\d\mathsf{C}_t^{\venus}\ra
=[1+\sigma\bar{n}]\d t.
\end{eqnarray}
Repeating this for other pair products of
$\d\mathsf{B}_t^\sharp$, $\d\tilde{\mathsf{B}}_t^\sharp$ and
$\d t$, multiplication rules for these increments can be
summarized in the following table:
\begin{eqnarray}
\begin{array}{l|ccccc}
&\d\mathsf{B}_t&\d\mathsf{B}_t^\dag&\d\tilde{\mathsf{B}}_t&\d\tilde{\mathsf{B}}_t^\dag&\d t\\\hline
\d\mathsf{B}_t&0&\phantom{\sigma\tau}[1+\sigma\bar{n}]\d t&\tau\bar{n}\d t&0&0\\
\d\mathsf{B}_t^\dag&\phantom{\sigma\tau}\bar{n}\d t&0&0&\tau[1+\sigma\bar{n}]\d t&0\\
\d\tilde{\mathsf{B}}_t&\sigma\tau\bar{n}\d t&0&0&\phantom{\tau}[1+\sigma\bar{n}]\d t&0\\
\d\tilde{\mathsf{B}}_t^\dag&0&\sigma\tau[1+\sigma\bar{n}]\d t&\phantom{\tau}\bar{n}\d t&0&0\\
\d t&0&0&0&0&0\\
\end{array}\;\;
\end{eqnarray}

\section{Treatment of fermions in Thermo Field Dynamics}
\label{phase}

We are deciding the double tilde conjugation rule and the thermal
state conditions for fermions \cite{Ojima81,Hayashi02} by considering
the system consisting of a vector field and Fad\-de\-ev-Po\-pov
ghosts \cite{Faddeev67}.

In the case of pure Abelian gauge field within the Feynman gauge,
the system is specified by the Hamiltonian
$H_{\mathrm{vf}+\mathrm{gh}}=\Hvf+\Hgh$ defined on the total state
vector space $\mathcal{V}=\mathcal{V}\vf\otimes\mathcal{V}\gh$.
$\Hvf$ and $H\gh$ are, respectively, Hamiltonians for the vector field and ghosts
defined on the vector field sector $\mathcal{V}\vf$ and the ghost
sector $\mathcal{V}\gh$ given by
\bea
\Hvf=-\int\d^3k\;\veps(\vec{k})\g^{\mu\nu}a_\mu^\dag(\vec{k})a_\nu(\vec{k}),
\eea
with $\veps(\vec{k})=|\vec{k}|$ being the energy spectrum and
$\g^{\mu\nu}=\mathrm{diag}(1,-1,-1,-1)$, and by
\bea
\Hgh=-i\int\d^3k\;\veps(\vec{k})
\ls\bar{c}^\dag(\vec{k})c(\vec{k})-c^\dag(\vec{k})\bar{c}(\vec{k})\rs.
\eea
Here, $a^\dag_\mu(\vec{k})$ and $a_\mu(\vec{k})$ are,
respectively, creation and annihilation operators of the gauge
field of the mode $\vec{k}$ satisfying the canonical commutation
relations
\bea
[a_\mu(\vec{k}),a^\dag_\nu(\vec{q})]=-\g_{\mu\nu}\delta^3(\vec{k}-\vec{q}),
\label{cr-vf1}
\eea
while $c^\dag(\vec{k})$ and $c(\vec{k})$ [$\bar{c}^\dag(\vec{k})$
and $\bar{c}(\vec{k})$] are, respectively, creation and
annihilation operators of ghosts [anti-ghosts] satisfying the
following canonical anti-commutation relations:
\bea
[c(\vec{k}),\bar{c}^\dag(\vec{q})]_+=
-[\bar{c}(\vec{k}),c^\dag(\vec{q})]_+=i\delta^3(\vec{k}-\vec{q}).
\eea
Other combinations of ghost/anti-ghost operators anti-commute with
each other.
The BRS charge -- generator of the BRS transformation \cite{Becchi76},
and the ghost charge \cite{Kugo78,Kugo79} acting on the total space
$\mathcal{V}$ are, respectively, given by
\bea
Q_B&=&-\int\d^3k\;k^\mu\ls a_\mu(\vec{k})c^\dag(\vec{k})+
a^\dag_\mu(\vec{k})c(\vec{k})\rs,\label{qb}\\
Q_c&=&\int\d^3k\ls c^\dag(\vec{k})\bar{c}(\vec{k})+
\bar{c}^\dag(\vec{k})c(\vec{k})\rs,\label{qc}
\eea
which satisfy
\bea
[iQ_c,Q_B]=Q_B.\label{qc-qb-commutator}
\eea

Let us introduce a set of new operators
$\{a_{(\sigma)}(\vec{k})\vert\sigma=+,-,\mathrm{L},\mathrm{S}\}$
through the relation

\bea
a_\mu(\vec{k})=a_{(\sigma)}(\vec{k})\epsilon_\mu^{(\sigma)}(\vec{k}),
\eea
where $\epsilon_\mu^{(\sigma)}(\vec{k})$ are polarization vectors
defined by
%
\bea
\epsilon_\mu^{(\pm)}(\vec{k})&=&(0,\vec{e}_\pm),\\
\epsilon^{(\mathrm{L})}_\mu(\vec{k})&=&-ik_\mu=-i(|\vec{k}|,\vec{k}),
\label{pv0}\\
\epsilon^{(\mathrm{S})}_\mu(\vec{k})&=&-i\bar{k}_\mu/2|\vec{k}|^2=
-i(|\vec{k}|,-\vec{k})/2|\vec{k}|^2,
\label{pv}
\eea
%
with $\vec{e}_\pm$ satisfying $\vec{e}_\pm\cdot\vec{k}=0$,
$\vec{e}_\pm^{\;*}\cdot\vec{e}_\mp=0$ and
$\vec{e}_\pm^{\;*}\cdot\vec{e}_\pm=1$. The polarization vectors
$\epsilon_\mu^{(\pm)}(\vec{k})$ correspond, respectively, to the
transverse modes with helicity $\pm1$, while
$\epsilon_\mu^{(\mathrm{L})}(\vec{k})$ and
$\epsilon_\mu^{(\mathrm{S})}(\vec{k})$ indicate, respectively, the
longitudinal mode and the scalar mode. With the definition
(\ref{pv0}) and (\ref{pv}), we see that
$\epsilon_\mu^{(\mathrm{L})*}(\vec{k})\cdot\epsilon^{(\mathrm{L}),\mu}(\vec{k})=
\epsilon_\mu^{(\mathrm{S})*}(\vec{k})\cdot\epsilon^{(\mathrm{S}),\mu}(\vec{k})=0$,
and
$\epsilon_\mu^{(\mathrm{L})*}(\vec{k})\cdot\epsilon^{(\mathrm{S}),\mu}(\vec{k})=1$.
Introducing a ``metric''
\bea
\g^{(\sigma\tau)}&=&\g_{(\sigma\tau)}=
\lp
\ba{rrrr}
-1&0&0&0\\
0&-1&0&0\\
0&0&0&\phantom{-}1\\
0&0&\phantom{-}1&0\\
\ea
\rp,
\label{metric}
\eea
we can define ``contravariant'' polarization vectors thro\-ugh
$\epsilon_{(\sigma)}^\mu(\vec{k})=\g_{(\sigma\tau)}\epsilon^{(\tau),\mu}(\vec{k})$,
and see that
%
\bea
\epsilon_{(\sigma)}^{\mu*}(\vec{k})\cdot\epsilon^{(\sigma),\nu}(\vec{k})&=&\g^{\mu\nu},\\
\epsilon^{(\sigma)*}_\mu(\vec{k})\cdot\epsilon^{(\tau),\mu}(\vec{k})&=&\g^{(\sigma\tau)}.
\eea
%
The commutation relations (\ref{cr-vf1}) being rewritten in terms
of operators $a_{(\sigma)}(\vec{k})$ and $a_{(\sigma)}^\dag(\vec{k})$ become
\bea
[a_{(\sigma)}(\vec{k}),a_{(\tau)}^\dag(\vec{q})]=
-\g_{(\sigma\tau)}\delta(\vec{k}-\vec{q}).
\label{cr-vf2}
\eea
Also the Hamiltonian for the vector field and generator of the BRS
transformation read
%
\bea
\Hvf&=&-\int\d^3k\;\veps(\vec{k})\g^{(\sigma\tau)}
a_{(\sigma)}^\dag(\vec{k})a_{(\tau)}(\vec{k}),\\
Q_B&=&-i\int\d^3k\ls a_{(\mathrm{S})}^\dag(\vec{k})c(\vec{k})-
a_{(\mathrm{S})}(\vec{k})c^\dag(\vec{k})\rs.\qquad
\eea
%

In the local covariant operator formalism \cite{Kugo78,Kugo79} of
gauge theories, the space $\mathcal{V}$ of state vectors has
inevitably an indefinite metric as can be seen by (\ref{cr-vf2})
with (\ref{metric}). The physical subspace
$\mathcal{V}_{\mathrm{phys}}$ of $\mathcal{V}$, defined by
\cite{Kugo79}
\bea
Q_{B}\mathcal{V}_{\mathrm{phys}}=0,
\label{subs-cond}
\eea
can be shown to have a positive semi-definite metric
\cite{Kugo79}. Dividing $\mathcal{V}_{\mathrm{phys}}$ by its
subspace $\mathcal{V}_0$ consisting of normless states, we have,
as a quotient space, the physical Hilbert space
$H_{\mathrm{phys}}$ ($=\mathcal{V}_{\mathrm{phys}}/\mathcal{V}_0$)
with positive definite metric in which the probabilistic
interpretation of quantum theory works. $H_{\mathrm{phys}}$ is
isomorphic to the Hilbert space $\mathscr{H}_{\rm phys}$ spanned
by the Fock states created by the cyclic operation of
$a^\dag_{(\pm)}(\vec{k})$ on a certain vacuum. The space spanned
by the Fock states created by $a^\dag_{(\mathrm{S})}(\vec{k})$ is
classified in $\mathcal{V}_{0}$, while the space spanned by the
Fock states created by $a^\dag_{(\mathrm{L})}(\vec{k})$ is
classified in a space complemented to
$\mathcal{V}_{\mathrm{phys}}$. This reflects the fact that
physical modes for photons are two transverse modes only.

As it is sufficient to pay attention to one mode in the following manipulation,
we will pick up a mode, say $\vec{k}$, from each type of particles, and drop
the index $\vec{k}$, for simplicity.
Let us span the state vector space $\mathcal{V}$ by means of a set
of the bases
$\{|\{n_{(\sigma)}\})\cdot|n_c,n_{\bar{c}})\}$
whose elements, being the bases of the vector field sector
$\mathcal{V}\vf$ and the ghost sector $\mathcal{V}\gh$,
respectively, are defined by
%
%
%
\bea
|\{n_{(\sigma)}\})&=&\prod_{\sigma=\pm,\rm{L},\rm{S}}\frac1{\sqrt{n_{(\sigma)}!}}
\lp a_{(\sigma)}^\dag\rp^{n_{(\sigma)}}|\{0\}),\quad\\
|n_c,n_{\bar{c}})&=&\lp c^\dag\rp^{n_c}\lp\bar{c}^\dag\rp^{n_{\bar{c}}}|0,0).
\label{bases}
\eea
%
%
%
They constitute the eigenstates of $\Hvf$, $\Hgh$
and $iQ_c$:
%
%
%
\bea
\Hvf|\{n_{(\sigma)}\})&=&E\vf(\{n_{(\sigma)}\})|\{n_{(\sigma)}\}),\\
\Hgh|n_c,n_{\bar{c}})&=&E\gh(n_c,n_{\bar{c}})
|n_c,n_{\bar{c}}),\\
iQ_c|n_c,n_{\bar{c}})&=&N\gh(n_c,n_{\bar{c}})
|n_c,n_{\bar{c}}),
\eea
%
%
%
with
$E\vf(\{n_{(\sigma)}\})=\veps\sum\nolimits_{\sigma}n_{(\sigma)}$,
$E\gh(n_c,n_{\bar{c}})=\veps\lp n_c+n_{\bar{c}}\rp$ and
$N\gh(n_c,n_{\bar{c}})=\lp n_c-n_{\bar{c}}\rp$, where
$n_{(\sigma)}$ are non-negative integers, and $n_c$ and
$n_{\bar{c}}$ take values of 0~or~1. We will denote the basis
vectors $\{|\{n_{(\sigma)}\})\cdot|n_c,n_{\bar{c}})\}$ by $\{|n)\}$
for brevity. Then, the metric tensor of $\mathcal{V}$ is
\bea
\eta_{n,m}=(n|m).
\eea

At the finite temperature, the statistical average of an
observable quantity $A$, satisfying
\bea
[Q_B,A]=0,
\eea
is given by \cite{Hata80}
\bea
\la A\ra
&=&\Tr A\rho P^{(0)}=\Tr A\rho \e^{\pi Q_c}\label{sa}
\eea
with $P^{(0)}$ being a projection operator onto
$\mathscr{H}_{\mathrm{phys}}$ and $\rho=Z^{-1}\e^{-\beta
H_{\mathrm{vf+gh}}}$ being the statistical operator acting on
$\mathcal{V}$ with the partition function $Z=\Tr\e^{-\beta
H_{\mathrm{vf+gh}}+\pi Q_c}$; the trace operation is taken in the space
$\mathcal{V}$. Here, for the second equality in
(\ref{sa}), we used the BRS-invariance of the statistical operator
\bea
[Q_B,\rho]=0.\label{rho-brs-inv}
\eea

Let us express the statistical average (\ref{sa}) as the
vacuum expectation in the doubled state space (thermal space)
$\hat{\mathcal{V}}=\mathcal{V}\otimes\tilde{\mathcal{V}}$ which is
introduced as follows. If $A$ is an operator on $\mathcal{V}$ so
that
\bea
A=\sum_{n,m}A_{nm}|n)(m|,
\eea
the corresponding vector $|A\ra$ in $\hat{\mathcal{V}}$ is
obtained as
\bea
|A\ra=A_{nm}|n,\tilde{m}\ra,
\eea
where
$\{|n,\tilde{m}\ra\equiv|\{n_{(\sigma)}\},\{\tilde{m}_{(\sigma)}\}\ra\cdot
|n_c,n_{\bar{c}},\tilde{m}_c,\tilde{m}_{\bar{c}}\ra\}$ is the set
of the bases spanning $\hat{\mathcal{V}}$ and defined through the
``principle of correspondence'' \cite{Arimitsu87a}:
%
%
%
\bea
|\{n_{(\sigma)}\},\{\tilde{m}_{(\sigma)}\}\ra&\longleftrightarrow&
|\{n_{(\sigma)}\})(\{m_{(\sigma)}\}|,\label{pc-vf}\\
|n_c,n_{\bar{c}},\tilde{m}_c,\tilde{m}_{\bar{c}}\ra&\longleftrightarrow&
|n_c,n_{\bar{c}})(m_c,m_{\bar{c}}|.\label{pc-gh}
\eea
%
%
%
The inner product in $\hat{\mathcal{V}}$ is given by
\bea
\la A|B\ra=\Tr A^\dag B.
\eea
Annihilation and creation operators acting on $\hat{\mathcal{V}}$
are defined through
%
%
%
\bea
\lp\ba{l}a_{(\tau)}\\a_{(\tau)}^\dag\ea\rp
|\{n_{(\sigma)}\},\{\tilde{m}_{(\sigma)}\}\ra
\leftrightarrow
\lp\ba{l}a_{(\tau)}\\a_{(\tau)}^\dag\ea\rp
|\{n_{(\sigma)}\})(\{m_{(\sigma)}\}|,
\eea
\bea
\lp\ba{l}\tilde{a}_{(\tau)}\\\tilde{a}_{(\tau)}^\dag\ea\rp
|\{n_{(\sigma)}\},\{\tilde{m}_{(\sigma)}\}\ra
\leftrightarrow
|\{n_{(\sigma)}\})(\{m_{(\sigma)}\}|
\lp\ba{l}a_{(\tau)}^\dag\\a_{(\tau)}\ea\rp,
\eea
%
%
%
for vector field, and through
%
\bea
\left(\begin{array}{c}
c\\
\bar{c}
\end{array}\right)
|n_c,n_{\bar{c}},\tilde{m}_c,\tilde{m}_{\bar{c}}\ra
\leftrightarrow
\left(\begin{array}{c}
c\\
\bar{c}
\end{array}\right)
|n_c,n_{\bar{c}})(m_c,m_{\bar{c}}|,
\eea
\bea
\left(\begin{array}{c}
c^\dag\\
\bar{c}^\dag
\end{array}\right)
|n_c,n_{\bar{c}},\tilde{m}_c,\tilde{m}_{\bar{c}}\ra
\leftrightarrow
\left(\begin{array}{c}
c^\dag\\
\bar{c}^\dag
\end{array}\right)
|n_c,n_{\bar{c}})(m_c,m_{\bar{c}}|,
\eea
\bea
\left(\begin{array}{c}
\tilde{c}\\
\tilde{\bar{c}}
\end{array}\right)
|n_c,n_{\bar{c}},\tilde{m}_c,\tilde{m}_{\bar{c}}\ra
\leftrightarrow
(-1)^{u+1}
|n_c,n_{\bar{c}})(m_c,m_{\bar{c}}|\!
\left(\begin{array}{c}
c^\dag\\
\bar{c}^\dag
\end{array}\right)\!,
\eea
\bea
\left(\begin{array}{c}
\tilde{c}^\dag\\
\tilde{\bar{c}}{}^\dag
\end{array}\right)
|n_c,n_{\bar{c}},\tilde{m}_c,\tilde{m}_{\bar{c}}\ra
\leftrightarrow
(-1)^u
|n_c,n_{\bar{c}})(m_c,m_{\bar{c}}|
\left(\begin{array}{c}
c\\
\bar{c}
\end{array}\right),
\eea
%
%
%
for ghosts, where
$u=N\gh(n_{c},n_{\bar{c}})-N\gh(m_{c},m_{\bar{c}})$. Also bases
$|\{n_{(\sigma)}\},\{\tilde{m}_{(\sigma)}\}\ra$ and
$|n_c,n_{\bar{c}},\tilde{m}_c,\tilde{m}_{\bar{c}}\ra$ are
generated from the vacuums $|\{0\},\{\tilde{0}\}\ra$ and
$|0,0,\tilde{0},\tilde{0}\ra$, respectively, as
%
\bea
|\{n_{(\sigma)}\},\{\tilde{m}_{(\sigma)}\}\ra=\prod_{\sigma}
\frac{(a_{(\sigma)}^\dag)^{n_{(\sigma)}}
(\tilde{a}_{(\sigma)}^\dag)^{m_{(\sigma)}}}
{\sqrt{n_{(\sigma)}!m_{(\sigma)}!}}|\{0\},\{\tilde{0}\}\ra,
\label{b-vf}\\
|n_c,n_{\bar{c}},\tilde{m}_c,\tilde{m}_{\bar{c}}\ra=(-1)^{v}(c^\dag)^{n_c}(\bar{c}^\dag)^{n_{\bar{c}}}
(\tilde{c}^\dag)^{m_c}(\tilde{\bar{c}}{}^\dag)^{m_{\bar{c}}}
|0,0,\tilde{0},\tilde{0}\ra,\qquad\label{b-gh}
\eea
%
where $v=m_cm_{\bar{c}}$. For the total vacuum, we will use a
collective designation $|0,\tilde{0}\ra$.

Let us introduce thermal vacuums $\la\theta|$ and
$|\mathit{0}(\beta)\ra$ $\in\hat{\mathcal{V}}$ such that
\bea
\la A\ra&=&\la\theta|A|\mathit{0}(\beta)\ra.
\eea
We require them to satisfy
\bea
\label{qbhatm-inv}
\la\theta|\hat{Q}_B^-=0,\quad
\hat{Q}_B^-|\mathit{0}(\beta)\ra=0,
\eea
and
\bea
\label{qchat-inv}
\la\theta|\hat{Q}_c=0,\quad
\hat{Q}_c|\mathit{0}(\beta)\ra=0,
\eea
where $\hat{Q}_B^-$ and $\hat{Q}_c$ are the generator of the BRS transformation
and the ghost hat-charge, respectively,~in~$\hat{\mathcal{V}}$~\cite{Ojima81}
\bea
\hat{Q}_B^-&=&Q_B-\tilde{Q}_B
=-i\ls a_{(\mathrm{S})}^\dag c-a_{(\mathrm{S})}c^\dag
+\tilde{a}_{(\mathrm{S})}^\dag\tilde{c}
-\tilde{a}_{(\mathrm{S})}\tilde{c}^\dag\rs,\label{qbhatm}\\
\hat{Q}_c&=&Q_c-\tilde{Q}_c
=c^\dag\bar{c}+\bar{c}^\dag c
-\tilde{c}^\dag\tilde{\bar{c}}-\tilde{\bar{c}}{}^\dag\tilde{c}.
\eea
To satisfy (\ref{qbhatm-inv}), we need a trick. Namely, by
rewriting (\ref{sa}) as
\bea
\la A\ra&=&\Tr\theta A\rho\e^{\pi Q_c}\theta^{-1},\label{trick}
\eea
we introduce an operator $\theta$ with the basic requirement that
its inverse exists. Then we settle the correspondence
\bea
\label{pc}
\la\theta|\leftrightarrow\theta,\quad
|\mathit{0}(\beta)\ra\leftrightarrow\rho\e^{\pi Q_c}\theta^{-1}.
\eea
It gives
%
\bea
\la\theta|\hat{Q}_c&=&\la\theta|\lp c^\dag\bar{c}+\bar{c}^\dag c
-\tilde{c}^\dag\tilde{\bar{c}}-\tilde{\bar{c}}{}^\dag\tilde{c}\rp\nonumber\\
&\leftrightarrow&\theta\lp c^\dag\bar{c}+\bar{c}^\dag c\rp
-\lp c^\dag\bar{c}+\bar{c}^\dag c\rp\theta
\nonumber\\
&=&[\theta,Q_c].\label{bra-qc-inv2}
\eea
With the requirement given by the first equality in
(\ref{qchat-inv}), the expression in (\ref{bra-qc-inv2}) is equal
to zero and tells us that $\theta$ and $Q_c$ commute with each
other, i.e. $[\theta,Q_c]=0$. Then
\bea
\hat{Q}_c|\mathit{0}(\beta)\ra&=&\lp c^\dag\bar{c}+\bar{c}^\dag c
-\tilde{c}^\dag\tilde{\bar{c}}-\tilde{\bar{c}}{}^\dag\tilde{c}\rp
|\mathit{0}(\beta)\ra\nonumber\\
&\leftrightarrow&
(c^\dag\bar{c}+\bar{c}^\dag c)\rho\e^{\pi Q_c}\theta^{-1}-
\rho\e^{\pi Q_c}\theta^{-1}(c^\dag\bar{c}+\bar{c}^\dag c)\nonumber\\
&=&[Q_c,\rho]\e^{\pi Q_c}\theta^{-1},\label{ket-qc-inv2}
\eea
%
and the second equality in (\ref{qchat-inv}) is automatically
satisfied as far as $[Q_c,\rho]=0$.

Based upon $[\theta,Q_c]=0$ and existence of $\theta^{-1}$,
let us try the following form of $\theta$:
\bea
\theta=\e^{i\phi_1(ic^\dag\bar{c})+i\phi_2(-i\bar{c}^\dag c)
+i\phi_3(ic^\dag\bar{c})(-i\bar{c}^\dag c)},
\label{theta1}
\eea
where, $\phi_1$, $\phi_2$ and $\phi_3$ are real numbers which
should be determined. Then, taking into account the first
correspondence in (\ref{pc}), the definition (\ref{qbhatm}) and
(\ref{theta1}), the calculation of the first equality in
(\ref{qbhatm-inv}) goes as
\bea
\la\theta|\hat{Q}_B^-&=&-i\la\theta|
\lp a_{(\mathrm{S})}^\dag c-a_{(\mathrm{S})}c^\dag
+\tilde{a}_{(\mathrm{S})}^\dag\tilde{c}
-\tilde{a}_{(\mathrm{S})}\tilde{c}^\dag\rp\nonumber\\
&\leftrightarrow&-i\lc\theta\lp a_{(\mathrm{S})}^\dag c-a_{(\mathrm{S})}c^\dag\rp
+\lp a_{(\mathrm{S})}c^\dag+a_{(\mathrm{S})}^\dag c\rp\theta\rc\nonumber\\
&=&-i\theta\lc a_{(\mathrm{S})}^\dag c-a_{(\mathrm{S})}c^\dag
+\e^{-i\phi_1-i\phi_3(-i\bar{c}^\dag c)}a_{(\mathrm{S})}c^\dag\right.\nonumber\\
&&\left.+\e^{i\phi_2+i\phi_3(ic^\dag\bar{c})}a_{(\mathrm{S})}^\dag c\rc.
\label{qb-bra}
\eea
If we take $\phi_1=0$, $\phi_2=\pi$ and $\phi_3=0$, we have
$\la\theta|\hat{Q}_B^-=0$ with the choice
\bea
\theta=\exp\{i\pi(-i\bar{c}^\dag c)\}.\label{theta2}
\eea
This structure for $\theta$ allows us to calculate the second
equality in (\ref{qbhatm-inv}) as
\bea
\hat{Q}_B^-|\mathit{0}(\beta)\ra&=&
-i\lp a_{(\mathrm{S})}^\dag c-a_{(\mathrm{S})}c^\dag
+\tilde{a}_{(\mathrm{S})}^\dag\tilde{c}
-\tilde{a}_{(\mathrm{S})}\tilde{c}^\dag\rp
|\mathit{0}(\beta)\ra\nonumber\\
&\leftrightarrow&-i\lc\lp a_{(\mathrm{S})}^\dag c-a_{(\mathrm{S})}c^\dag\rp
\rho\e^{\pi Q_c}\theta^{-1}{}
+\rho\e^{\pi Q_c}\theta^{-1}
\lp-a_{(\mathrm{S})}c^\dag-a_{(\mathrm{S})}^\dag c\rp\rc
\nonumber\\
&=&Q_B\rho\e^{\pi Q_c}\theta^{-1}+
\rho\e^{\pi Q_c}Q_B\theta^{-1}\nonumber\\
&=&[Q_B,\rho]\e^{\pi Q_c}\theta^{-1},
\label{qb-ket}
\eea
where we also used $[\e^{\pi Q_c},Q_B]_+=0$ which is obtained from
(\ref{qc-qb-commutator}). Taking into account the BRS-invariance
of the statistical operator $\rho$ (\ref{rho-brs-inv}), the
expression (\ref{qb-ket}) is equal to zero, and both requirements
(\ref{qbhatm-inv}) are fulfilled with the choice (\ref{theta2}).
We see that introduction of factors $\theta$ and $\theta^{-1}$
indeed is necessary to satisfy the BRS-invariance of the thermal
vacuums $\la\theta|$ and $|\mathit{0}(\beta)\ra$. Expression of
the unit operator in $\mathcal{V}$
\bea
\textbf{1}&=&\sum_{n,m}|n)\eta_{n,m}^{-1}(m|,
\eea
and correspondences (\ref{pc}) with (\ref{theta2}) enable us to
see the structure of thermal vacuums as
\bea
\la\theta|&=&\sum_{n,m}(\eta_{n,m}^{-1})^*\la n,\tilde{m}|\theta\nonumber\\
&=&\sum_{n_{(\sigma)},m_{(\sigma)}}(\eta_{n_{(\sigma)},m_{(\sigma)}}^{-1})^*
\la\{n_{(\sigma)}\},\{\tilde{m}_{(\sigma)}\}|
{}\nonumber\\
&&\times
\mathop{\sum_{n_c,n_{\bar{c}}}}\limits_{m_c,m_{\bar{c}}}\relax
(\eta_{(n_c,n_{\bar{c}}),(m_c,m_{\bar{c}})}^{-1})^*
\la n_c,n_{\bar{c}},\tilde{m}_c,\tilde{m}_{\bar{c}}|\theta\nonumber\\
&=&\la\{0\},\{\tilde0\}|\exp\lc-\g^{(\sigma\tau)}\tilde{a}_{(\sigma)}a_{(\tau)}\rc
\la0,0,\tilde0,\tilde0|[1+i\tilde{c}\bar{c}][1+i\tilde{\bar{c}}c]\nonumber\\
&=&\la0,\tilde0|\exp\lc i\tilde{c}\bar{c}+i\tilde{\bar{c}}c
-\g^{(\sigma\tau)}\tilde{a}_{(\sigma)}a_{(\tau)}\rc,\label{c46}
\eea
and
\bea
|\mathit{0}(\beta)\ra&=&Z^{-1}\exp
\lc-\e^{-\beta\veps}\g^{(\sigma\tau)}a_{(\sigma)}^\dag\tilde{a}_{(\tau)}^\dag
- i\e^{-\beta\veps}\left(c^\dag\tilde{\bar{c}}{}^\dag
+\bar{c}^\dag\tilde{c}^\dag \right) \rc
|0,\tilde0\ra.
\eea
It may be instructive to note that considering
\bea
\label{qbhatp-inv}
\la\theta|\hat{Q}_B^+=0,\quad
\hat{Q}_B^+|\mathit{0}(\beta)\ra=0,
\eea
with $\hat{Q}_B^+=Q_B+\tilde{Q}_B$, instead of (\ref{qbhatm-inv})
with (\ref{qbhatm}), leads to the choice
$\theta=\exp\{i\pi(ic^\dag\bar{c})\}$.

After determination of parameters $\phi_i$ the thermal state
conditions with the ghost operators can be derived through the
following steps. First, for the bra-vacuum we see
\bea
\la\theta|
\lp\ba{r}c^\dag\\\bar{c}^\dag\ea\rp
\longleftrightarrow&\theta
\lp\ba{r}c^\dag\\\bar{c}^\dag\ea\rp\nonumber\\
&\|\\
\la\theta|
\lp\ba{r}\tilde{c}\\-\tilde{\bar{c}}\ea\rp
\longleftrightarrow&
\lp\ba{r}c^\dag\\-\bar{c}^\dag\ea\rp\theta,\nonumber
\eea
therefore
\bea
\la\theta|
\lp\ba{r}c^\dag\\\bar{c}^\dag\ea\rp
&=&
\la\theta|
\lp\ba{r}\tilde{c}\\-\tilde{\bar{c}}\ea\rp.
\eea
Similarly, taking into account structures of the statistical
operator $\rho$, the ghost charge $Q_c$, (\ref{qc}), and $\theta$,
(\ref{theta2}), for the ket-vacuum we have
\bea
\lp\ba{r}c\\\bar{c}\ea\rp|\mathit{0}(\beta)\ra
&\longleftrightarrow&\lp\ba{r}c\\\bar{c}\ea\rp
\rho\;\e^{\pi Q_c}\theta^{-1}\nonumber\\
&&\qquad\qquad\|\\
\lp\ba{r}\e^{-\beta\veps}\tilde{c}^\dag\\-\e^{-\beta\veps}\tilde{\bar{c}}{}^\dag\ea\rp
|\mathit{0}(\beta)\ra
&\longleftrightarrow&
\rho\;\e^{\pi Q_c}\theta^{-1}
\lp\ba{r}\e^{-\beta\veps}c\\-\e^{-\beta\veps}\bar{c}\ea\rp,
\nonumber
\eea
which gives
\bea
\lp\ba{r}c\\\bar{c}\ea\rp|\mathit{0}(\beta)\ra
&=&
\e^{-\beta\veps}\lp\ba{r}\tilde{c}^\dag\\-
\tilde{\bar{c}}{}^\dag\ea\rp
|\mathit{0}(\beta)\ra.
\eea

The double tilde conjugation rule must be defined so that it
leaves thermal vacuums unchanged. To this end, we put
%
\bea
\lp\tilde{a}_{(\sigma)}\rp^\sim&=&a_{(\sigma)},\\
\lp\ba{c}\tilde{c}\\\tilde{\bar{c}}\ea\rp^\sim&=&
\lp\ba{c}\xi{c}\\\bar\xi{\bar{c}}\ea\rp,
\eea
%
and determine parameters $\xi$ and $\bar\xi$ so that
$\la\theta|^\sim=\la\theta|$ and
$|\mathit0(\beta)\ra^\sim=|\mathit0(\beta)\ra$. Taking the tilde
conjugation of (\ref{c46}) we have
\bea
\la\theta|^\sim
&=&\la0,\tilde{0}|^\sim
\exp\{-i\xi c\tilde{\bar{c}}-i\bar{\xi}\bar{c}\tilde{c}-\g^{(\sigma\tau)}
a_{(\sigma)}\tilde{a}_{(\tau)}\}\nonumber\\
&=&\la0,\tilde{0}|
\exp\{i\xi\tilde{\bar{c}}c+i\bar{\xi}\tilde{c}\bar{c}-\g^{(\sigma\tau)}
\tilde{a}_{(\sigma)}a_{(\tau)}\},
\eea
where we assumed $\la0,\tilde0|^\sim=\la0,\tilde0|$. The requirement
$\la\theta|^\sim=\la\theta|$ gives $\xi=1$ and $\bar\xi=1$, which leads to
\bea
\lp\ba{c}\tilde{c}\\\tilde{\bar{c}}\ea\rp^\sim=
\lp\ba{c}{c}\\{\bar{c}}\ea\rp.\label{DTC}
\eea
As a consequence, we obtain the tilde-invariance of the thermal
ket-vacuum too:
\bea
|\mathit{0}(\beta)\ra^\sim&=&\lp Z^{-1}
\exp\{-\e^{-\beta\veps}\g^{(\sigma\tau)}a^\dag_{(\sigma)}\tilde{a}^\dag_{(\tau)}
-i\e^{-\beta\veps}(c^\dag\tilde{\bar{c}}^\dag+\bar{c}^\dag\tilde{c}^\dag)\}
|0,\tilde{0}\ra\rp^\sim\nonumber\\
&=&Z^{-1}\exp\{-\e^{-\beta\veps}\g^{(\sigma\tau)}\tilde{a}^\dag_{(\sigma)}a^\dag_{(\tau)}
+i\e^{-\beta\veps}(\tilde{c}^\dag\bar{c}^\dag+\tilde{\bar{c}}^\dag c^\dag)\}
|0,\tilde{0}\ra\nonumber\\
&=&|\mathit{0}(\beta)\ra.
\eea

A similar line of reasoning can be used to derive the tilde
conjugation rule and the thermal state conditions for a system
consisting of physical fermions. Let us consider the system
specified by the Hamiltonian
\bea
H\pf=\int\d^3k\;\veps(\vec{k})\,a^\dag(\vec{k})a(\vec{k})
\eea
with $a(\vec{k})$ and $a^\dag(\vec{k})$ being, respectively,
fermion annihilation and creation operators satisfying the
canonical anti-commutation relation
\bea
[a(\vec{k}),a^\dag(\vec{q})]_+=\delta(\vec{k}-\vec{q}).
\eea
As in previous consideration, in the following manipulation we
will pay attention to one mode, say $\vec{k}$, and drop the index
$\vec{k}$ for simplicity. In that case the bases of the state vector space
$\mathcal{V}\pf$ will be denoted as $|0)$ and $|1)$ defined by
$a|0)=0$ and $|1)=a^\dag|0)$.

At the finite temperature, the statistical average of an
observable quantity $A$ is given by
\bea
\la A\ra=\Tr A\rho\pf\label{sapf}
\eea
with $\rho\pf$ being the statistical operator of the system
\bea
\rho\pf=Z\pf^{-1}\e^{-\beta H\pf}
=Z\pf^{-1}\left[|0)(0|+\e^{-\beta\veps}|1)(1|\right],\label{rho-rf}
\eea
where $Z\pf$ is the partition function.

Within the doubled state space
$\hat{\mathcal{V}}\pf=\mathcal{V}\pf\otimes\tilde{\mathcal{V}}\pf$,
the statistical average (\ref{sapf}) can be expressed in terms of
the vacuum expectation with respect to the thermal bra- and
ket-vacuums for physical fermions. The bases of
$\hat{\mathcal{V}}\pf$ are defined through the principle of
correspondence:
\bea
\label{pcpf}
|n,\tilde{m}\ra&\longleftrightarrow&|n)(m|,
\eea
where $n$ and $m$ take values of $0$ or $1$. Annihilation and
creation operators acting on $\hat{\mathcal{V}}\pf$ are
defined through
%
\bea
\lp\ba{l}a\\a^\dag\ea\rp
|n,\tilde{m}\ra&\longleftrightarrow&
\lp\ba{l}a\\a^\dag\ea\rp
|n)(m|,\\
\tilde{a}|n,\tilde{m}\ra&\longleftrightarrow&(-1)^{n-m+1}|n)(m|a^\dag,\\
\tilde{a}^\dag|n,\tilde{m}\ra&\longleftrightarrow&(-1)^{n-m}|n)(m|a.
\eea
%
The bases $|n,\tilde{m}\ra$ are generated from the vacuum
$|0,\tilde{0}\ra$:
\bea
|n,\tilde{m}\ra=(a^\dag)^n(\tilde{a}^\dag)^m|0,\tilde{0}\ra.
\eea

This time we don't have peculiar symmetries for thermal vacuums
like in the case of gauge theories with the BRS symmetry. However,
in the derivation of thermal vacuums, let us use a trick similar
to the one in (\ref{trick})
\bea
\la A\ra=\Tr\,\e^{i\phi a^\dag a}A\rho\,\e^{-i\phi a^\dag a}
=\la\theta|A|\mathit{0}(\beta)\ra,\label{vepf}
\eea
where $\phi$ is a real number which should be decided. We settle
the correspondence for $\la\theta|$ and $|\mathit{0}(\beta)\ra$ as
\bea
\la\theta|\longleftrightarrow\e^{i\phi a^\dag a},\quad
|\mathit{0}(\beta)\ra\longleftrightarrow\rho\pf\e^{-i\phi a^\dag a}.
\eea
They are normalized, $\la\theta|\mathit{0}(\beta)\ra=1$, and
generated from $|0,\tilde{0}\ra$ as
%
\bea
\la\theta|&=&\la0,\tilde{0}|\ls1+\e^{i\phi}\tilde{a}a\rs,\label{bra-rf-1}\\
|\mathit{0}(\beta)\ra
&=&Z\pf^{-1}\ls1+\e^{-i\phi}\e^{-\beta\veps}a^\dag\tilde{a}^\dag\rs
|0,\tilde{0}\ra.\label{ket-rf-1} \eea
%
A requirement of the tilde invariance for the thermal bra-vacuum
\bea
\la\theta|^\sim=\la\theta|
\eea
determines the tilde conjugation rule for physical fer\-mi\-on
operators up to the phase factor
\bea
\label{dtc-rf}
(a)^\sim=\tilde{a},\quad
(\tilde{a})^\sim=-\e^{2i\phi}a.
\eea
We have seen in (\ref{DTC}) that the ghost operators, which are
\textit{fermion} operators, were unchanged under the double tilde
conjugation. Let us adopt the same rule for physical fermion
operators, i.e. put $\phi=\pi/2$ to obtain
\bea
\label{tcpf}
(a)^\sim=\tilde{a},\quad
(\tilde{a})^\sim=a.
\eea
With this choice of $\phi$, the thermal vacuums for physical
fermions read
%
\bea
\la\theta|&=&\la0,\tilde{0}|\ls1+i\tilde{a}a\rs,\label{bra-rf}\\
|\mathit{0}(\beta)\ra
&=&Z\pf^{-1}\ls1-i\e^{-\beta\veps}a^\dag\tilde{a}^\dag\rs
|0,\tilde{0}\ra\label{ket-rf}
\eea
%
and satisfy the following thermal state conditions
\bea
\la\theta|\tad=-i\la\theta|a,\quad
\ta|\mathit{0}(\beta)\ra=i\e^{-\beta\veps}\ad|\mathit{0}(\beta)\ra.
\eea

\section{Correlation of random force operators}
\label{crf}

The random force operators are of the Wiener process whose first
and second moments are given by real $c$-numbers:
%
\bea
\la\d F_t\ra&=&\la\d F^\dag_t\ra=0,\label{dFt-average-a}\\
\la\d F_t\d F_t\ra&=&\la\d F^\dag_t\d F^\dag_t\ra=0,\label{dFt-average-b}\\
\la\d F_t\d F^\dag_t\ra&=&\mbox{a real $c$-number},\\
\la\d F^\dag_t\d F_t\ra&=&\mbox{a real $c$-number},
\eea
%
where $\la\cdots\ra=\la|\cdots|\ra$ represents the random average
referring to the random force operators $\d F_t$. From
(\ref{dFt-average-a}), (\ref{dFt-average-b}) and TSC
(\ref{TSC-Ft}) we have for operators (\ref{sde.11}) and (\ref{dWt})
%
\bea
\la\d W_t\ra=\la\d\tilde{W}_t\ra&=&\la\d W_t^{\venus}\ra=\la\d\tilde{W}_t^{\venus}\ra=0,\\
\la\d W_t\d W_s\ra&=&\la\d\tilde{W}_t\d\tilde{W}_s\ra=0,
\label{cor-1}
\eea
%
while from (\ref{bra-w-venus-zero}) it follows
%
\bea
\la\d W_t^{\venus}\d W_s\ra&=&\la\d W_t^{\venus}\d\tilde{W}_s\ra=0,\\
\la\d W_t^{\venus}\d W_s^{\venus}\ra&=&\la\d W_t^{\venus}\d\tilde{W}_s^{\venus}\ra=0,
\label{cor-2}
\eea
%
and their tilde conjugates. Using (\ref{cor-1}) to (\ref{cor-2})
the explicit structure of $\d\hat{M}_t\d\hat{M}_t$ in
(\ref{H-St-a}) is written as
\bea
\d\hat{M}_t\d\hat{M}_t=-2\sigma\d W_t\d\tilde{W}_t
\alpha^{\venus}\tilde{\alpha}^{\venus}
+\lambda\lp\d W_t\d W_t^{\venus}\alpha^{\venus}\alpha
+\d\tilde{W}_t\d\tilde{W}_t^{\venus}\tilde{\alpha}^{\venus}\tilde{\alpha}\rp
\label{dMdM}
\eea
in a ``weak sense''%
\footnote{In the case of classical systems it
corresponds to the stochastic convergence.}. 
We demand that the
Stratonovich type time evolution generator should not contain a
diffusion term, i.e. the term proportional to
$\alpha^{\venus}\tilde\alpha^{\venus}$. Then the correlation
$\la\d W_t\d\tilde{W}_t\ra$ is determined to be
\bea
\d W_t\d\tilde{W}_t=\tau\lc2\kappa(t)[n(t)+\eta]+\dot{n}(t)\rc\d t
=\la\d W_t\d\tilde{W}_t\ra\label{crf.5}
\eea
so that $\hat{\mPi}_{\mathrm{D}}$ in (\ref{H-St-a}) is cancelled
by the first term in the r.h.s.\ of (\ref{dMdM}). Here, the first
equality in (\ref{crf.5}) should be understood in a weak sense as
well. Expression (\ref{crf.5}) is compatible with the assumption
that the process is white. Let us put the subscript $F$ to
$\Sigma^<(t)$ in the Boltzmann equation in order to remember that
it is due to the interaction with the random force $\d F_t$:
\bea
\dot{n}(t)&=&-2\kappa(t)n(t)+i\Sigma^<_F(t).\label{crf.6}
\eea
Making use of two previous equations, we have
\bea
i\Sigma^<_F(t)\d t&=&2\kappa(t)n(t)\d t+\dot{n}(t)\d t\nonumber\\
&=&-2\kappa(t)\eta\d t+\sigma\tau\la\d W_t\d\tilde{W}_t\ra\nonumber\\
&=&-2\kappa(t)\eta\d t+\la\d F^\dag_t\d F_t\ra
+\nu\ls\la\d F_t\d F^\dag_t\ra-\sigma\la\d F^\dag_t\d F_t\ra\rs,
\label{crf.7}
\eea
where (\ref{sde.11}) has been used, and $\mu$ has been erased with
the help of (\ref{ief.26}).

We can assume that the quantity $\eta$ may depend on $\nu$, i.e.
$\eta=\eta(\nu)$, and that the physical quantities $\kappa(t)$,
$\Sigma^<_F(t)$, $\la\d F^\dag_t\d F_t\ra$, and $\la\d F_t\d
F^\dag_t\ra$ may not depend on $\nu$. Then, differentiating
equation (\ref{crf.7}) with respect to $\nu$, one has
\bea
0&=&-2\kappa(t)\frac{\p\eta}{\p\nu}\d t+
\la\d F_t\d F^\dag_t\ra-\sigma\la\d F^\dag_t\d F_t\ra.\quad
\label{b8}
\eea
This leads to
%
\bea
\frac{\p\eta}{\p\nu}&=&k(t),
\eea
which is solved as
\bea
\eta&=&k(t)\nu+l(t),
\eea
%
where $k(t)$ and $l(t)$ are real numbers independent of $\nu$.
With this solution one has
\bea
\la\d F_t\d F^\dag_t\ra-\sigma\la\d F^\dag_t\d F_t\ra&=&2\kappa(t)k(t)\d t,
\label{FFk}
\eea
and
\bea
i\Sigma^<_F(t)\d t&=&-2\kappa(t)l(t)\d t+\la\d F^\dag_t\d F_t\ra,
\label{crf.11}
\eea
which leads to
%
\bea
\la\d F^\dag_t\d F_t\ra
&=&\lc2\kappa(t)[l(t)+n(t)]+\dot{n}(t)\rc\d t,\label{crf.12a}
\label{crf.12}
\eea
where we have used (\ref{crf.6}).
The substitution of (\ref{crf.12a}) into (\ref{FFk}) gives us
\bea
\la\d F_t\d F^\dag_t\ra
=\lc2\kappa(t)[k(t)+\sigma l(t)+\sigma n(t)]+\sigma\dot{n}(t)\rc\d t.
\label{crf.12b}
\eea
%

For the case of stationary quantum stochastic process, the
Boltzmann equation (\ref{crf.6}) reduces to
\bea
\dot{n}(t)&=&-2\kappa[n(t)-\bar{n}],\label{crf.13}
\eea
where $\bar{n}$ is the average quantum number in equilibrium.
Therefore, (\ref{crf.12}) and (\ref{crf.12b}) reduce, respectively, to
%
\bea
\la\d F^\dag_t\d F_t\ra&=&2\kappa[\bar{n}+l(t)]\d t,\\
\la\d F_t\d F^\dag_t\ra&=&2\kappa[k(t)+\sigma l(t)+\sigma\bar{n}]\d t.
\label{crf.14}
\eea
%
Since in the white noise assumption the Boltzmann equation
(\ref{crf.13}) is compatible with the stationary process specified by
\cite{Haken70}
%
\bea
\la\d F^\dag_t\d F_t\ra&=&2\kappa\bar{n}\d t,\\
\la\d F_t\d F^\dag_t\ra&=&2\kappa[1+\sigma\bar{n}]\d t,
\label{corr-stationary}
\eea
%
one concludes now that
\bea
l(t)=0,&\quad&k(t)=1,
\label{crf.16}
\eea
which leads to
\bea
\eta=\nu,&\quad&\xi=\mu.
\label{crf.17}
\eea
Note that the result (\ref{corr-stationary}) can be obtained using the
Bogoliubov transformation as it is described in Appendix~\ref{bfbm}.

Substituting (\ref{crf.16}) into (\ref{crf.12}) and (\ref{crf.12b}), one obtains
%
\bea
\la\d F^\dag_t\d F_t\ra&=&\ls2\kappa(t)n(t)+\dot{n}(t)\rs\d t,\\
\la\d F_t\d F^\dag_t\ra&=&\lc2\kappa(t)[1+\sigma n(t)]+\sigma\dot{n}(t)\rc\d t,
\quad
\label{sde.12}
\eea
%
which leads to
\bea
\la\d W_t\d W_t^{\venus}\ra=\la\d F_t,\d F_t^\dag\ra
-\sigma\la\d F_t^\dag,\d F_t\ra=2\kappa(t)\d t.
\label{cor-3}
\eea
Assembling (\ref{dMdM}), (\ref{crf.5}), (\ref{crf.17}) and
(\ref{cor-3}) one obtains expression (\ref{GFDT}).


\end{document}